\documentclass[]{jfm}

\usepackage{graphicx}
\usepackage{newtxtext}
\usepackage{newtxmath}
\usepackage{natbib}
\usepackage{hyperref}
\hypersetup{
    colorlinks = true,
    urlcolor   = blue,
    citecolor  = black,
}

\newcommand{\RomanNumeralCaps}[1]
\linenumbers

\usepackage[utf8]{inputenc} 
\usepackage[T1]{fontenc}
\usepackage{amsmath,amssymb}
\usepackage{mathrsfs}
\usepackage{pstricks}
\usepackage{psfrag}
\usepackage{graphicx}
\usepackage{mathtools}
\usepackage{float}
\usepackage{xparse}
\usepackage{xspace}
\usepackage{physics,bm}
\usepackage{textcomp}
\usepackage{ulem}
\usepackage{textcomp}
\usepackage{stmaryrd}
\usepackage{array}
\usepackage{xcolor}
\usepackage{url}
\usepackage{enumitem}
\usepackage{accents}

\newcommand{\balpha}{\bm{\alpha}}
\newcommand{\btheta}{\bm{\theta}}
\newcommand{\bbeta}{\bm{\beta}}

\newcommand{\ubold}{\bm{u}}
\newcommand{\vbold}{\bm{v}}
\newcommand{\bu}{\ubold}
\newcommand{\bv}{\vbold}
\newcommand{\bx}{\bm{x}}
\newcommand{\by}{\bm{y}}

\newcommand{\xd}{\{x\rbrace}
\newcommand{\ud}{\{u\rbrace}
\newcommand{\vd}{\{v\rbrace}

\newcommand{\bq}{\bm{q}}

\newcommand{\bT}{\bm{\tau}}
\newcommand{\bn}{\boldsymbol{n}}
\providecommand\bcdot{\boldsymbol{\cdot}}
\providecommand\bnabla{\boldsymbol{\nabla}}
\newcommand{\mbf}{\boldsymbol}

\newcommand{\cdd}[1]{#1}
\newcommand{\review}[1]{#1}

\graphicspath{{figures/}}

\title{\review{Data-driven kinematics-consistent model order reduction of fluid-structure interaction problems: application to deformable microcapsules in a Stokes flow}}

\author{Claire Dupont\aff{1}, Florian De Vuyst\aff{2} \and Anne-Virginie Salsac\aff{1}\corresp{\email{anne-virginie.salsac@utc.fr}}}

\affiliation{\aff{1}Biomechanics and Bioengineering Laboratory (UMR 7338), 
Universit{\'e} de Technologie de Compi{\`e}gne - CNRS, 60203 Compi{\`e}gne, France
\aff{2} Laboratory of Applied Mathematics of Compiègne,  Universit{\'e} de Technologie de Compi{\`e}gne, CS 60319, 60203 Compi{\`e}gne, France}

\begin{document}
\maketitle

\begin{abstract}
\review{In this paper, we present a generic approach of a dynamical data-driven model order reduction technique for three-dimensional fluid-structure interaction problems. 
A low-order continuous linear differential system is identified from snapshot solutions of a high-fidelity solver. 
The reduced order model (ROM)  uses different ingredients like proper orthogonal decomposition (POD), dynamic mode decomposition (DMD) and Tikhonov-based robust identification techniques. An interpolation method is used to predict the capsule dynamics for any value of the governing non-dimensional parameters that are not in the training database.
Then a dynamical system is built from the predicted solution.
Numerical evidence shows the ability of the reduced model to predict the time-evolution of the capsule deformation from its initial state, whatever the  parameter values.  Accuracy and stability properties of the resulting low-order dynamical system are analyzed numerically. The numerical experiments show a very good agreement, measured in terms of modified Hausdorff distance between capsule solutions of the full-order and low-order models both in the case of confined and unconfined flows.  
This work is a first milestone to move towards real time simulation of fluid-structure problems, which can be extended to non-linear low-order systems to account for strong material and flow non-linearities. It is a valuable innovation tool for rapid design and for the development of innovative devices.}
\end{abstract}


\section{Introduction}

\review{Fluid-structure interaction (FSI) problems often occur in  Engineering (aircraft and automotive industries, wind turbines) as well as in medical applications (cardiovascular systems, artificial organs, artificial valves, medical devices, etc.). Today the design of such systems usually requires advanced studies and high-fidelity (HF) numerical simulations become an essential tool of computed-aided analysis. However, computational FSI is known to be very time-consuming even on high-performance computing facilities. Usually, engineering problems are parameterized and the search of suitable designs require numerous computer experiments leading to prohibitive computational times. 
For particular applications such as the tracking of drug carrier capsules flowing in blood vessels, it would be ideal to have real-time simulations for a better understanding of the behaviour of the dynamics and for efficiency assessment. Unfortunately, today high-fidelity real-time FSI simulations are far from being reached with current High Performance Computing (HPC) facilities. 

A current trend is to use machine learning (ML) or artificial intelligence (AI) tools such as artificial neural networks (ANN). Such tools learn numerical simulations from HF solvers and try to map entry parameters with output criteria in an efficient way, with response times far less than HF ones, say 3 or 4 orders of magnitude smaller. In some sense, heavy HF computations and training stage are done in an offline stage, and learned ANNs can be used online for real time evaluations and analysis. However, ML and ANN today are not fully satisfactory for dynamical problems, and/or the training stage itself may be time consuming, thus requiring more Central Processing Unit (CPU) time.
Another option is the use of model order reduction (MOR). Reduced-order modeling (ROM) can be seen as a 'grey-box' supervized ML methodology, taking advantage of the expected low-order dimensionality of the FSI mechanical problem. By 'grey-box' we mean that the low-dimensional encoding of the ML process is based on mechanical principles and a man-made preliminary dimensionality reduction study. This allows one for a better control of the ROM accuracy and behaviour. 
There are two families of MOR: intrusive and non-intrusive approaches. The intrusive approaches use physical equations. The low-order model is derived by setting the physical problem on a suitable low-dimensional space. The accuracy can be very good, but the price to pay is the generation of a new code which can be a tedious and long task. The non-intrusive approach does not require heavy code development. It is based on HF simulation results used as entry data. Although it is not based on high-fidelity physical equations, a non-intrusive approach can include a priori physical informations, like e.g. meaningful physical features, prototype of system of equations, pre-computed principal components, consistency with physical principles, etc.

In the recent literature, efficient intrusive ROMs for FSI
have been proposed e.g. in \citep{Quarteroni2016}. But to our knowledge
there are far less contributions in non-intrusive ROMs dedicated to FSI.

In this paper, we propose a data-driven model order reduction approach for FSI problem which is consistent with the equations of kinematics and is designed from a low-order meaningful system of equations.
As case of study, we focus on the motion of a microcapsule, a droplet surrounded by a membrane, subjected to a confined and unconfined Stokes flow.  

Artificial microcapsules can be used in various industrial applications such as in cosmetics \citep{Miyazawa2000, Casanova2016}, food industry \citep{Yun2021} and biotechnology, where drug targeting is a high potential application \citep{Ma2013,Abuhamdan2021, Ghiman2022}.
Once in suspension in an external fluid, capsules are subjected to hydrodynamics forces, which may lead to large membrane deformation, wrinkle formation or damage. 
The numerical model must be able to capture the time-evolution of the nonlinear 3D large deformations of the capsule membrane. 
Different numerical strategies are possible to solve the resulting large systems of equations \citep{Lefebvre2007, Hu2012, Ye2017, Tran2020}. However, they all have long computational times.

Different approaches have been used over the past decade to accelerate the computations, such as HPC (e.g. \cite{Zhao2010}) and Graphics Processing units (e.g. \cite{Matsunaga2014}).
More recently, reduced order models have been proposed to predict the motion of capsules suspended in an external fluid flow.
In~\citet{Quesada2021}, the authors used the large amount of data generated by numerical simulations to show how relevant it is to recycle these data to produce lower-dimensional problem using physics-based reduced order models. However, their method can only  predict the steady-state capsule deformed shape. 
\citet{Boubehziz2021} show for the first time the efficiency of data-driven model-order reduction technique to predict the dynamics of the capsule in a microchannel. However, the method is cumbersome as it requires two POD bases, one to predict the velocity field, the other to capture the shape evolution over time. And then they reconstruct the solution in the parameter space thanks to diffuse approximation (DA) strategy.

The proposed method serves different objectives.
We have designed the  method to be non intrusive for practical uses of  existing high-fidelity FSI solver (also referred to as the Full-Order Model, or FOM). That means that the ROM methodology should be data-driven.
We also want the ROM to be consistent with the equations of kinematics. The model must thus return the displacement  $\{u\}$ and velocity $\{v\}$ fields from a few snapshots provided by the FOM. 
It must otherwise be able to predict the solution for any parameter vector in predefined admissible domain.
Finally, the kinematics-consistent data-driven reduced-order model of capsule dynamics must ideally open the way to real-time simulations. 
To do so, we use a coupling between Proper Orthogonal Decomposition (POD) and Dynamic Mode Decomposition (DMD), as well as a Tikhonov regularization for robustness purposes and interpolation to predict solutions for any parameter value.

As indicated above, we mainly consider the case of an initially spherical capsule flowing in a microfluidic channel with a square cross-section.
The corresponding FOM was developed by \citet{Hu2012} and used to get a complete numerical database of the three-dimensional capsule dynamics as a function of the parameters of the problem: the capsule-to-tube confinement ratio, hereafter referred to as size ratio $a/\ell$ and the capillary number $Ca$, which measures the ratio between the viscous forces acting onto the capsule membrane and the membrane elastic forces. 
For clarity reasons, different ROMs are introduced  with increasing levels of generality, as detailed in Table~\ref{tab:1}. First, we consider a fixed parameter vector, and get a space-time ROM in the form of a low-order dynamical system.  Next, we generate such $N$ ROMs for the~$N$ parameter samples that fill the admissible parameter domain, and then assess the uniform accuracy (space-time accuracy over the whole sample set). Finally, we propose a strategy to derive a general space-time-parameter ROM for any value of the parameter vector $(Ca,a/\ell)$ in the admissible space.

\begin{table}
    \centering
    \begin{tabular}{|c|c|c|c|}\hline
        Nb of parameter & ROM output type & Verification & Related Section(s)  \\ 
        samples for data && (accuracy) & in the paper\\ \hline\hline
       1  & 1 space-time ROM & Space-time accuracy & Sections \ref{sec:ROM} and  \ref{sec:1example} \\ \hline
       $N$ & $N$ space-time ROM & Uniform space-time  & Section~\ref{sec:database} \\ 
       && accuracy on the & \\ 
       && sample set & \\ \hline
       $N$ & 1 space-time-parameter ROM & Uniform accuracy & Section~\ref{sec:interpolation} \\
       & (any parameter couple) && \\ \hline
    \end{tabular}
    \caption{\review{General methodology, stepwise procedure for ROM construction of increasing level of generality.}}
    \label{tab:1}
\end{table}

The paper is organized as follows. First, we present the physics of the problem and the FOM in Section \ref{sec:FOM}. 
The  strategy used to develop a non-intrusive space-time ROM is detailed in Section \ref{sec:ROM}. The results are presented for a given configuration in Section \ref{sec:1example}. The accuracy is then estimated in Section \ref{sec:database} on the entire database, formed by all the cases that have reached a stationary state. Finally in Section \ref{sec:interpolation} we present the methodology of space-time-parameter ROM. The ROM accuracy is confirmed by some numerical experiments.}

\section{Full-order microcapsule model, parameters and quantities of interest}
\label{sec:FOM}

\subsection{Problem description}

\begin{figure}
\centering
\includegraphics[width=0.9\textwidth]{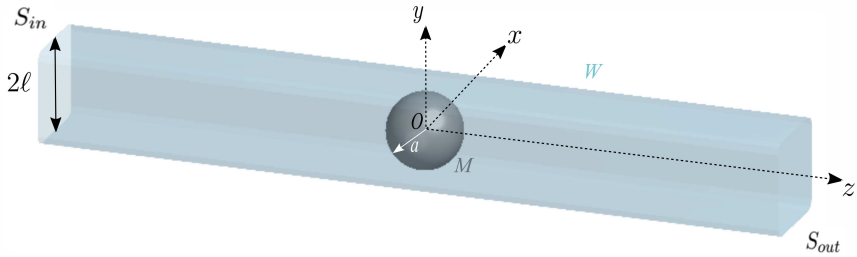}
\caption{Sketch of the model geometry showing an initially spherical capsule of radius $a$ placed in a channel  with a constant square section of side $2\ell$. }
\label{IllustrationPb} 
\end{figure}

An initially spherical capsule of radius $a$ flows within a long microfluidic channel having a constant square section of side $2\ell$ (Figure \ref{IllustrationPb}). The suspending fluid and capsule liquid core are incompressible Newtonian fluids with the same kinematic viscosity $\eta$.  

The capsule liquid core is enclosed by a hyperelastic isotropic membrane. Its thickness is assumed to be negligible compared to the capsule dimension. The membrane is thus modeled as a surface devoid of bending stiffness with surface shear modulus $G_S$. 
The two non-dimensional governing parameters of the problem are the size ratio $a/\ell$ and the capillary number 
\begin{equation}
Ca= \eta V/G_S
\end{equation}
where $V$ is the mean axial velocity of the undisturbed external Poiseuille flow.\\

The flow Reynolds number is assumed to be very small. We solve the Stokes equations in the external ($\beta = 1$) and internal fluids ($\beta = 2$), together with the membrane equilibrium equation to determine the dynamics of the deformable capsule within the microchannel. 

For the fluid problem, we denote $\bv^{(\beta)}$, $\boldsymbol{\sigma}^{(\beta)}$ and $p^{(\beta)}$ the velocity, stress and pressure fields in the two fluids. These parameters are non-dimensionalized using $\ell$ as characteristic length, $\ell/V$ as characteristic time and $G_S \ell$ as characteristic force. 
The non-dimensional Stokes equations 
 \begin{equation}
\nabla p^{(\beta)} = Ca \nabla^2\bv^{(\beta)},\quad\nabla\cdot\bv^{(\beta)}=0,\quad\beta=1,2
\label{eq:Stokes}
\end{equation}
are solved in the domain bounded by the cross sections $S_{in}$ at the tube entrance and $S_{out}$ at the exit. These cross sections are assumed to be both located far from the capsule. The reference frame $(O,\bm{x}, \bm{y}, \bm{z})$ is centered at each time step on the capsule center of mass $O$ in the high-fidelity code, but the displacement of the capsule center of mass along the tube axis~$\bm{Oz}$ is computed.

The boundary conditions of the problem are the following ones:

\begin{itemize}
\item~The velocity field is assumed to be the unperturbed flow field  on $S_{in}$ and $S_{out}$, i.e. the flow disturbance vanish far from the capsule.
\item~The pressure is uniform on $S_{in}$ and $S_{out}$.
\item~A no-slip boundary condition is assumed at the channel wall $W$ and on the capsule membrane $M$:
\begin{equation}
\forall \bx\in W, \bv(\bx)=\bm{0};\quad 
\forall \bx \in M,\ \bv(\bx)=\frac{\partial \bu}{\partial t}.
\label{eq:noslipcondition}
\end{equation}
\item~The normal load $\bm{n}$ on the capsule membrane $M$ is continuous, i.e. the non-dimensionalized external load per unit area $\bq$ exerted by both fluids  is due to the viscous traction jump:
\begin{equation} \label{eq:traction_jump}
(\boldsymbol{\sigma}^{(1)}-\boldsymbol{\sigma}^{(2)})\bcdot\bn =  \bq
\end{equation} where $\bn$ is the unit normal vector pointing towards the suspending fluid. 
\end{itemize}

To close the problem, the external load $\bq$ on the membrane is deduced from the local equilibrium equation, which, in absence of internia, can be written as
\begin{equation}
  \bnabla_s \bcdot \boldsymbol \tau + \bq = \mbf 0
  \label{eq:equi-mb}
\end{equation}
where $\bT$ is the non-dimensionalized Cauchy tension tensor (forces per unit arclength in the deformed plane of the membrane) and $\bnabla_s \bcdot$ is the surface divergence operator. We assume that the membrane deformation is governed by the strain-softening neo-Hookean law. The principal Cauchy tensions can then be expressed as
\begin{equation}
\tau_1 = \frac{G_S}{\lambda_1 \lambda_2}\left[ \lambda_1^2 - \frac{
1}{(\lambda_1 \lambda_2)^2}\right] ~(\text{likewise for } \tau_2),
\end{equation}
where $\lambda_1$ and $\lambda_2$ are the principal extension ratios measuring the in-plane deformation.

\subsection{Numerical procedure}

The FSI problem is solved by coupling a finite element method that determines the capsule membrane mechanics with a boundary integral method that solves for the fluid flows \citep{Walter2010,Hu2012}. Thanks to the latter, only the boundaries of the flow domain, i.e the channel entrance $S_{in}$ and exit $S_{out}$, the channel wall and the capsule membrane have to be discretized to solve the problem.
The mesh of the initially spherical capsule is generated by subdividing the faces of the icosahedron (regular polyhedron with 20 triangular faces) inscribed in the sphere until reaching the desired number of triangular elements. At the last step, nodes are added at the middle of all the element edges to obtain a capsule mesh with 1280 $P_2$ triangular elements and 2562 nodes, which correspond to a characteristic mesh size $\Delta h_C= 0.075\,a$. The channel mesh of the entrance surface $S_{in}$ and exit surface $S_{out}$ and of the channel wall is generated using 
\texttt{Modulef} (INRIA, France). The central portion of the channel, where the capsule is located, is refined. The channel mesh comprises 3768 $P_1$~triangular elements and 1905~nodes.

At time $t=0$, a spherical capsule is positioned with its center of mass $O$ on the channel axis. 
At each time step, the in-plane stretch ratio $\lambda_1$ and $\lambda_2$ are computed from the nodes deformation.
 The elastic tension tensor $\boldsymbol \tau$ is then deduced from the values of $\lambda_1$ and $\lambda_2$. The finite element method is used to solve the weak form of the membrane equilibrium equation~\eqref{eq:equi-mb} and determine the external load $\bq$.

Applying the boundary integral method, the velocity of the nodes on the capsule membrane reads \citep{Pozrikidis1992}:
\begin{equation}
\bv(\bx) = \bv^\infty(\bx) - \frac{1}{8\pi\mu_F} \left[ \int_M \bm{J}(\bm{r})\cdot\bq dS(\by) + \int_W \bm{J}(\bm{r})\cdot \bm{f} dS(\by) -\Delta P \int_{S_{out}} \bm{J}(\bm{r})\cdot \bm{n}\, dS(\by)\right]
\label{eq:flo1}
\end{equation}
for any $\bx$ in the spatial domain when the suspending and internal fluids have the same viscosity. The vector $\bm{f}$ is the disturbance wall friction due to the capsule, $\Delta P$ is the additional pressure drop and $\bm{r}=\by-\bx$. 

To update the position of the membrane nodes, the nodal displacement $\bu$ is computed by integrating equation \eqref{eq:noslipcondition} in time.
The procedure is repeated until the desired non-dimensional time $VT/\ell$.

For later development, it is more convenient to work on the condensed abstract form of the system. The full order semi-discrete FSI system to solve consists of the kinematics and the membrane equilibrium algebraic equations:
\begin{align}
& \dot\ud = \vd, \qquad t\in [0,T],\label{eq:flo2}\\
& \vd = \varphi(\ud) \label{eq:flo3}
\end{align}
where $\varphi$ is a nonlinear mapping from $\mathbb{R}^{3d}$ to~$\mathbb{R}^{3d}$ and $d$ is the number of nodes on the membrane.
Regarding time discretization, a Runge-Kutta Ralston scheme is used:
\begin{align*}
    & \{\widehat u^{n+2/3}\} = \{u^n\} + \frac{2}{3}\Delta t \, \{v^{n}\},
    \\
    & \{\widehat v^{n+2/3}\} = \{\varphi\}(\{\widehat u^{n+2/3}\}), \\
    & \{u^{n+1}\} = \{u^n\} + \Delta t \, \left(\frac{1}{4}\{v^{n}\}+\frac{3}{4}\{\widehat v^{n+2/3}\}\right),\\
    & \{v^{n+1}\} = \{\varphi\}(\{u^{n+1}\}), \\
    & \{u^0\} = \{0\},\ \{v^{0}\} = \{\varphi\}(\{0\})
\end{align*}
%
%
\noindent where $\Delta t>0$ is a constant time step and $\ud^n$ and $\vd^n$ respectively represent the discrete membrane displacement field and the discrete membrane velocity field at discrete time~$t^n=n\Delta t$. The initial condition is simply $\ud^0=\lbrace 0\rbrace $.

\begin{figure}
\centering
\includegraphics[width=0.55\textwidth]{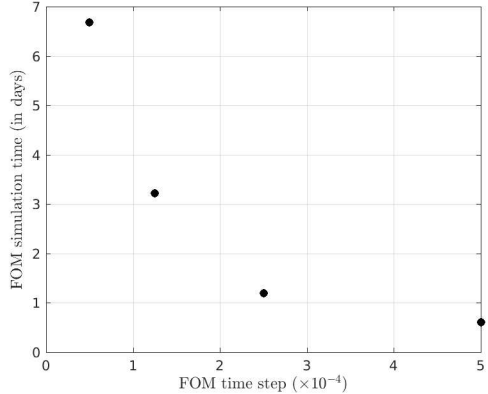}
\caption{Simulation time of the dynamics of the capsule over a non-dimensional time $Vt/\ell=10$ ($a/\ell=0.7$) according to the time step. Simulations were performed on a workstation equipped with 2  processors Intel\textsuperscript{\textregistered} Xeon\textsuperscript{\textregistered} Gold 6130 CPU (2.1 GHz).}
\label{SimuTime} 
\end{figure}

The whole numerical scheme is subject to some Courant-Friedrichs-Lewy (CFL) type stability condition on the time step \citep{Walter2010} because of its explicit nature.
The numerical method is conditionally stable if the time step $\Delta t$ satisfies
\begin{equation}
\frac{V}{\ell}\Delta t < O\left( \frac{\Delta h_C}{\ell}Ca\right).
\label{StabilityCondition}
\end{equation}
From the computational point of view, the resolution of~\eqref{eq:flo3} at each time step requires i) the computation of  the disturbance wall friction $\bm{f}$ at all the wall nodes, ii)  the additional pressure drop $\Delta P$, iii) the traction jump $\bq$ at the membrane nodes and iv) the boundary integrals for each node.
The resulting numerical FOM may thus be time-consuming, depending on the membrane discretization and the number of time steps. Figure \ref{SimuTime} shows that the evolution of the computational cost when $a/\ell=0.7$, considering the mesh discretization described above and a workstation equipped with 2  processors Intel\textsuperscript{\textregistered} Xeon\textsuperscript{\textregistered} Gold 6130 CPU (2.1 GHz). A week of computation is sometimes necessary to simulate the dynamics of an initially spherical capsule in a microchannel over the non-dimensional time $VT/\ell=10$.

For that reason, a model-order reduction (MOR) strategy is studied in this paper, in order to reduce the computational time by several orders of magnitude. ROMs try to approximate solutions of the initial problem by strongly lowering the dimensionality of the numerical model, generally using a reduced basis (RB) of suitable functions, then derive a low-order system of equations.

In the case of differential algebraic equations (DAE) like~\eqref{eq:flo2}-\eqref{eq:flo3}, the reduced system of equations to find should also be of DAE nature. Remark that it is often possible to reformulate DAEs as a system of ordinary differential equations (ODEs) \citep{Ascher1998}. 
In the next section, we give details on the chosen ROM methodology for the particular case and context of FSI capsule problem.  

\section{\review{Non-intrusive space-time model-order reduction strategy}}
\label{sec:ROM}

\review{In this section, the parameter couple $\btheta=(Ca,a/\ell)$ is fixed, thus we omit the dependency of the solutions with respect to $\btheta$ for the sake of simplicity}.
For the derivation of the ROM model, we consider the semi-discrete time-continuous version of the FOM, i.e.~\eqref{eq:flo2}-\eqref{eq:flo3}.
\subsection{Dimensionality reduction and reduced variables for displacements and velocities}
Assume first that, for any $t\in [0,T]$, the discrete velocity field can be accurately approximated according to the expansion
\begin{equation}
    \vd(t) \approx \sum_{k=1}^K \beta_k(t)\, \lbrace \phi_k\rbrace 
\label{eq:flo6}
\end{equation}

\noindent for some orthonormal modes $\lbrace \phi_k\rbrace \in\mathbb{R}^d$ and real coefficients $\beta_k(t)$.
The truncation rank $K\leq d$ is of course expected to be far less than $d$ as expected in a general ROM methodology. From the kinematics equations we have
\begin{align*}
\ud(t) &= \int_0^t \vd(s)\, ds\\
       &\approx \int_0^t \beta_k(s)\, \lbrace \phi_k\rbrace \ ds
\end{align*}
so that the displacement field can be accurately represented by 
\begin{equation}
\ud(t) \approx \sum_{k=1}^K\alpha_k(t)\, \lbrace \phi_k\rbrace 
\label{eq:flo7}
\end{equation}
where $\displaystyle{\alpha_k(t)=\int_0^t \beta_k(s)\, ds}$. The coefficients
$(\alpha_k(t))_k$ and $(\beta_k(t))_k$ are called the reduced variables. For the  sake of readability and mental correspondence between full-order unknowns and reduced ones, we will use the convenient notations
\[
\balpha(t)=(\alpha_1(t),...,\alpha_K(t))^T,\quad
\bbeta(t)=(\beta_1(t),...,\beta_K(t))^T
\]
where the exponent $^T$ denotes the transpose of the matrix.
The condensed matrix forms of~\eqref{eq:flo7} and~\eqref{eq:flo6}
respectively are
\begin{equation}
\ud(t) \approx Q\, \balpha(t),\quad
\vd(t) \approx Q\, \bbeta(t),
\end{equation}
where $Q=[\lbrace \phi_1\rbrace ,...,\lbrace \phi_K\rbrace ]\in\mathscr{M}_{dK}$. Since the modes $\lbrace \phi_k\rbrace $ are assumed to be orthonormal (for the standard Euclidean inner product), the matrix $Q$ is a semi-orthogonal matrix, i.e. $Q^TQ=I_K$. In particular, we have
$\balpha(t)\approx Q^T\, \ud(t)$ and $\bbeta(t)=Q^T\, \vd(t)$.

\review{Note that the modes $\lbrace \phi_k\rbrace $ and reduced variables $\balpha$, $\bbeta$  are determined for each parameter set ($Ca, a/\ell$), but a common value of the truncation rank $K$  is chosen for all the sets. Its practical computation will be detailed in a next subsection, as well as that of the modes $\lbrace \phi_k\rbrace $}.
\subsection{ROM prototype}
The expressions $\lbrace \tilde u\rbrace (t)=Q\, \balpha(t)$ and
$\lbrace \tilde v\rbrace (t)=Q\, \bbeta(t)$ provide low-order representations of displacement and velocity fields respectively. We can now write equations for the reduced vectors $\balpha(t)$ and $\bbeta(t)$ respectively. In this subsection, let us consider a projection Galerkin-type approach. Let us denote $\langle.,.\rangle$
the standard Euclidean scalar product in $\mathbb{R}^d$. Considering 
a test vector $\lbrace w\rbrace $ in $W=span(\lbrace \varphi_1\rbrace ,...,\lbrace \varphi_K\rbrace )$, we look for an approximate displacement field $\tilde \ud(t)$ solution of the projected kinematics equations
\[
\langle \frac{d}{dt}\lbrace \tilde u\rbrace (t),\lbrace w\rbrace \rangle = 
\langle \lbrace \tilde v\rbrace (t),\lbrace w\rbrace  \rangle
\quad \forall\, \lbrace w\rbrace \in W.
\]
By considering each test vector $\lbrace w\rbrace =\lbrace \varphi_k\rbrace $, we get the consistent reduced kinematics equation
\begin{equation}
\dot\balpha = \bbeta.
\label{eq:flo9}
\end{equation}
Consider now the projected field $\lbrace \tilde v\rbrace (t)$ which is solution of the
system of algebraic equations (Galerkin approach):
\begin{equation}
\langle \lbrace \tilde v\rbrace (t),\lbrace w\rbrace \rangle = \langle \varphi(\lbrace \tilde u\rbrace (t)),\lbrace w\rbrace  \rangle \quad \forall\, \lbrace w\rbrace \in W.
\label{eq:flo10}
\end{equation}
\review{
Again by taking the test vector $\{w\}=\{\phi_k\}$, we have
\[
\{\phi_k\}^T Q \beta(t) = \{\phi_k\}^T  \varphi(Q \balpha(t)).
\]
Considering all $k$ in $\{1,...,K\}$, since $Q=[\lbrace \phi_1\rbrace ,...,\lbrace \phi_K\rbrace ]$ and $Q^TQ=I_K$ we get}
\[
Q^T Q \bbeta(t) = \bbeta(t) = Q^T \varphi(Q \balpha(t)).
\]
It is in the form
\begin{equation}
\bbeta(t) = \varphi_r(\balpha(t))
\label{eq:flo11}
\end{equation}
with the mapping $\varphi_r:\mathbb{R}^K\rightarrow\mathbb{R}^K$
defined by $\varphi_r(\balpha)=Q^T\varphi(Q\balpha)$. 
We get a reduced-order algebraic equilibrium equation.
Unfortunately,
because of the nonlinearities in $\varphi$, the computation of
$\varphi_r(\balpha)$ requires high-dimensional $O(d)$ operations, making this approach irrelevant from the performance point of view.
A possible solution to deal with the nonlinear terms would be to use for example Empirical Interpolation Methods (EIM) \citep{Barrault2004} but from the algorithm and implementation point of view, this would lead to an intrusive approach with specific code developments. We here rather adopt a linearization strategy in the following sense: by derivating~\eqref{eq:flo11} with respect to time, we get
\[
\dot\bbeta(t) = \frac{\partial\varphi_r}{\partial \balpha}(\balpha(t))
\, \dot\balpha(t).
\]
Thanks to the reduced kinematics equation~\eqref{eq:flo9}, we get
\begin{equation}
\dot\bbeta(t) = \frac{\partial\varphi_r}{\partial \balpha}(\balpha(t))
\ \bbeta(t).
\label{eq:12}
\end{equation}
Since $\varphi_r$ is hard to evaluate, it is even  harder to evaluate its differential. But the differential $\frac{\partial\varphi_r}{\partial \balpha}(\balpha(t))$ can be approximated itself, or replaced by some matrix  $A(t)$.
Then we get a ROM structure (ROM prototype) in the form
\begin{align}
& \dot\balpha = \bbeta(t), \label{eq:flo13} \\
& \dot\bbeta(t) = A(t) \ \bbeta(t) \label{eq:flo14}.
\end{align}
The differential system~\eqref{eq:flo13}-\eqref{eq:flo14} is linear with variable coefficient matrix $A(t)\in \mathscr{M}_K(\mathbb{R})$. It can be written in matrix form
\begin{equation}
\frac{d}{dt}\begin{pmatrix} \balpha(t)\\ \bbeta(t) \end{pmatrix}
= \underbrace{\begin{pmatrix} [0] & I_K \\ [0] & A(t) \end{pmatrix}}_{=\mathbb{A}(t)}
\begin{pmatrix} \balpha(t)\\ \bbeta(t) \end{pmatrix}.
\label{eq:flo15}
\end{equation}
The spectral properties of the differential system~\eqref{eq:flo15} are related to the spectral properties of matrix $A(t)$. In particular, if all the (complex) eigenvalues $\lambda_k(t)$ of $A(t)$ are such that
$\Re(\lambda_k(t))<0$ for all~$k$ (uniformly distributed in time), then the system is dissipative.
\subsection{Nonintrusive approach, SVD decomposition and POD modes}
One of the requirements of this work is to achieve a non-intrusive reduced-order model, meaning that the effective implementation of the ROM does not involve tedious low-level code development into the FOM code. For that, a data-based approach is adopted: from the FOM code, it is possible to compute FOM solutions $(\ud^n,\vd^n)$ at discrete times $t^n$, $n=0,...,N$ ($t^N=N\Delta t=T$), then store some snapshot solutions (called snapshots) into a database for data analysis and later design of a ROM. Proper Orthogonal Decomposition (POD) \citep{Berkooz1993} is today a well-known dimensionality reduction approach to determine the principal components from solutions of partial differential equations. The Sirovich's snapshot variant approach \citep{Sirovich1987} is based on snapshot solutions from a FOM to get \textit{a posteriori} empirical POD modes $\lbrace \varphi_k\rbrace $. For the sake of simplicity, assume that the snapshot solutions are all the discrete FOM solution at simulation instants.
Applying a singular value decomposition (SVD) to the displacement  snapshot matrix
\[
\mathbb{S}^u = \left[\bu^1,\bu^2,...,\bu^N\right],
\]
of size $d\times N$, we get the SVD decomposition 
\begin{equation}
\mathbb{S}^u = U \Sigma V^T
\end{equation}
with orthogonal matrices $U\in\mathscr{M}_d(\mathbb{R})$,
$V\in\mathscr{M}_N(\mathbb{R})$ and the singular value matrix
$\Sigma=diag(\sigma_k)\in\mathscr{M}_{d\times N}(\mathbb{R})$, with $\sigma_k\geq 0$ for all $k$ organized in decreasing order:
$\sigma_1\geq \sigma_2\geq ...\geq\sigma_{\min(d,N)}\geq 0$.
From SVD theory, for a given accuracy threshold $\varepsilon>0$,
the truncation rank $K=K(\varepsilon)$ is computed as the smallest integer such that the inequality
\begin{equation}
 \frac{\displaystyle{\sum_{k=K+1}^{\min(d,N)} \sigma_k^2}}{\displaystyle{\sum_{k=1}^{\min(d,N)}\sigma_k^2}}\leq \varepsilon
\label{eq:RIC}
\end{equation}
holds \citep{Shawe2004}. Proceeding like that, it is shown that the relative orthogonal projection error of the snapshots~$\vd^n$ onto the linear subspace $W$ spanned by the $K$ first eigenvectors of $U$ is controlled by $\varepsilon$. Denoting~$\pi^W$ the linear orthogonal projection operator over $W$, we have:
\[
\sum_{n=1}^N \|\vd^n-\pi^W\vd^n\|^2 \leq \varepsilon
\sum_{n=1}^N \|\vd^n\|^2.
\]
The matrix $Q$ is obtained as the restriction of $U$ to its $K$ first columns. 

\subsection{Data-driven identification of coefficient matrix}\label{sec34}
%
The system~\eqref{eq:flo13}-\eqref{eq:flo14} is still not closed since the coefficient matrices $A(t)$ are unknowns.
From FOM data, one can try to identify the matrices by minimizing some residual function that measures the model discrepancy.
The simplest linear model corresponds to the case where $A(t)$ is searched as a time-constant matrix~$A$. In this case, equation~\eqref{eq:flo14} becomes
$\dot\bbeta(t) = A \, \bbeta(t)$. This is the scope of this article. From the time continuous problem, one could determine the matrix $A$ by minimizing the least square functional
\[
\min_{A\in\mathscr{M}_K(\mathbb{R})} \frac{1}{2}\int_0^T
\|\dot\bbeta(t)-A \bbeta(t)\|^2\, dt.
\]
But practically, we only have velocity snapshot data at discrete times and
we do not have access to the time derivatives of the velocity fields. 
So the following numerical procedure is adopted: from the velocity snapshot matrix 
$\mathbb{S}^v=[\vd^1,...,\vd^N]$, we compute first the reduced snapshots variables:
\[
\bbeta^n = Q\,  \vd^n,\quad n=1,...,N.
\]
Next, we determine a matrix $A$ that minimizes the least square cost function:
\begin{equation}
\min_{A\in\mathscr{M}_K(\mathbb{R})}\ \frac{1}{2} \sum_{n=1}^{N-1}
\left\|\frac{\bbeta^{n+1}-\bbeta^n}{\Delta t}-A\bbeta^n \right\|^2
\label{eq:flo17}
\end{equation}
In~\eqref{eq:flo17}, the finite difference $\dfrac{\bbeta^{n+1}-\bbeta^n}{\Delta t}$
is a first-order approximation (in $\Delta t$) of $\dot\bbeta$ at time $t^n$.
In appendix \ref{app:a}, we provide a mathematical analysis of the effect of time discretization in~\eqref{eq:flo17} about the impact on the stability of the resulting identified differential system compared to the initial one. 

The minimization problem~\eqref{eq:flo17} can be written in condensed matrix form
\begin{equation}
\min_{A\in\mathscr{M}_K(\mathbb{R})}\ \frac{1}{2} \|\mathbb{Y}-A\mathbb{X}\|_F^2
\label{eq:flo18}
\end{equation}
with the two data matrices
\begin{equation}
\mathbb{X} = \left[\bbeta^1,\bbeta^2,...,\bbeta^{N-1}\right],\quad
\mathbb{Y} = \left[\frac{\bbeta^2-\bbeta^1}{\Delta t},...,\frac{\bbeta^N-\bbeta^{N-1}}{\Delta t} \right].
\label{eq:flo19}
\end{equation}
Because $\mathbb{X}$ and $\mathbb{Y}$ store reduced variables (of size $K$), for a sufficient number of discrete snapshot times, these two matrices are
horizontal ones. We will assume that the rank of $\mathbb{X}$ is always maximal, i.e. equal to $K$.
The least-square solution $A$ of~\eqref{eq:flo18} is then given by
\begin{equation}
A = \mathbb{Y}\mathbb{X}^\dagger
\label{eq:flo20}
\end{equation}
where $\mathbb{X}^\dagger=\mathbb{X}^T(\mathbb{X}\mathbb{X}^T)^{-1}$ is the Moore-Penrose pseudoinverse matrix of $\mathbb{X}$.
This least square approach has close connections with SVD-based Dynamic Mode Decomposition (DMD) \citep{Schmid2010,Kutz2016}.
    \subsection{Tikhonov least-square regularized formulation}
    From standard spectral theory arguments, it is expected that the POD coefficients rapidly decay when $k$ increases as soon as both displacement and velocity fields are smooth enough. 
    A possible side effect is the bad condition number of the matrix $\mathbb{X}$, since the last rows of $\mathbb{X}$ have small coefficients (thus leading to row vectors close to zero 'at the scale' of the first row of $\mathbb{X}$). Even if the solution $A$ in~\eqref{eq:flo20} always exists, the solution may be sensitive to perturbations, noise or round-off errors.
    In order to get a robust identification approach, one can regularize the least-square problem~\eqref{eq:flo18} by adding a Tikhonov regularization term (see e.g. \citep{Aster2005})
    \begin{equation}
    \min_{A\in\mathscr{M}_K(\mathbb{R})}\ \frac{1}{2} \|\mathbb{Y}-A\mathbb{X}\|_F^2 + \frac{\mu}{2} \|\mathbb{X}\|_F^2\, \|A\|_F^2
    \label{eq:flo21}
    \end{equation}
    where the scalar $\mu>0$ is the regularization coefficient. The factor
    $\|\mathbb{X}\|_F^2$ in the regularization term has been added for scaling purposes.
    The solution $A_\mu$ of~\eqref{eq:flo21} is given by
    \begin{equation}
    A_\mu = \mathbb{Y} \mathbb{X}^T\left(\mathbb{X}\mathbb{X}^T+\mu \|\mathbb{X}\|_F^2\, I_K\right)^{-1}.
    \label{eq:flo22}
    \end{equation}
    \subsubsection*{Choice of optimal regularization coefficient}
    Of course, the solution matrix $A_\mu$ depends on the regularization coefficient $\mu$ and one can ask what is the optimal choice for $\mu$.
    There is a trade-off between the approximation quality measured by the residual $\|\mathbb{Y}-A_\mu\mathbb{X}\|_F$ and the norm solution $\|A_\mu\|_F$. The minimization of $\|A_\mu\|_F$ should ensure that unneeded features will not appear in the regularized solution.
    When plotted on the log-log scale, the curve of optimal values 
    $\mu\mapsto \|A_\mu\|_F$ versus the residual $\mu\mapsto\|\mathbb{Y}-A_\mu\mathbb{X}\|_F$ often takes on a characteristic L shape \citep{Aster2005}. A design of experiment with
    the test of 
    different values of $\mu$ (starting say from \review{$10^{-12}$ to $10^{-5}$})
    generally allow to find quasi-optimal values of $\mu$ located at the corner of the L-curve, thus providing a good trade-off between the two criteria.
    \subsection{Reduced-order continuous dynamical system}
    Once the matrix $A_\mu$ has been determined, we get the reduced-order continuous dynamical system
    \begin{align}
    & \dot\balpha = \bbeta, \label{eq:flo23}\\
    & \dot\bbeta = A_\mu\, \bbeta \label{eq:flo24}
    \end{align}
    with initial conditions $\balpha(0)=\bm{0}$, $\bv(0)=Q^T \varphi(\lbrace 0\rbrace )$. At any time $t$, one can go back to the high-dimensional physical space using the POD modes: $\ud(t)=Q\balpha(t)$, $\xd(t)=\lbrace X\rbrace +\ud(t)$, 
    $\vd(t)=Q \bbeta(t)$. As already mentioned, the system can be written in condensed matrix form
    \begin{equation}
    \dot{\bm{w}} = \mathbb{A}_\mu\, \bm{w}
    \label{eq:flo25}
    \end{equation}
    where $\bm{w}(t)=(\balpha(t),\bbeta(t))^T$ and
    $\mathbb{A}_\mu = \begin{pmatrix}[0]_K & I_K\\ [0]_K & A_\mu. \end{pmatrix}$.
    
    The exact analytical solution of~\eqref{eq:flo25} is
    \begin{equation}
    \bm{w}(t) = \exp(\mathbb{A}_\mu t)\,\bm{w}(0).
    \label{eq:flo26}
    \end{equation}
    The stability of the differential system depends on the spectral structure of $\mathbb{A}_\mu$, or equivalently on the spectrum of $A_\mu$. Because of the stability of the fluid-capsule coupled system and from accurate solutions of the FOM solver, one can hope that the solution $A_\mu$ of the least-square identification problem has the expected spectral properties. This will be studied and discussed in the numerical experimentation section. From the kinetic energy point of view, it is shown in appendix \ref{app:b} that the stability of the kinetic energy is linked to the property of the (real) spectrum of the symmetric matrix $(A_\mu+A_\mu^T)/2$. 
    \subsubsection*{Model consistency with steady states} 
    A steady state in our context is defined by a capsule that reaches a constant
    velocity $\vd_\infty$, so that the motion becomes a translation flow in time
    in the direction $\vd_\infty$. From~\eqref{eq:flo6}, this shows that 
    $\bbeta(t)$ also reaches a constant vector $\bbeta_\infty$, and
    $\dot\bbeta = 0$ at steady state. As a consequence, from~\eqref{eq:flo24},
    we get $A_\mu \bbeta_\infty=0$, meaning that $0$ is an eigenvalue of $A_\mu$ with $\bbeta_\infty$ as eigenvector. As a conclusion, the matrix $A_\mu$ must have zero
    in its spectrum in order to be consistent with the existence of steady states.
    \subsection{Reduced-order discrete dynamical system}
    Of course, it is also possible to derive a discrete dynamical system from the continuous one by using a standard time advance scheme. For example the explicit forward Euler scheme with a constant time step $\Delta t$  gives
    \begin{align}
    & \balpha^{n+1} = \balpha^n + \Delta t\, \bbeta^n,  \label{eq:flo27} \\
    & \bbeta^{n+1} = \bbeta^n + \Delta t\, A_\mu \bbeta^n. \label{eq:flo28}
    \end{align}
    By multiplying~\eqref{eq:flo27} by $Q$ we get the space-time approximate solution
    \[
    \ud^{n+1} = \ud^n + \Delta t\, \vd^n,
    \]
    so the ROM model is completely consistent with the kinematics equation.
    Stability properties of the discrete system are linked to the spectral properties of the matrix
    \[
    A_\mu^\Delta = \begin{pmatrix}
    I_K & \Delta t\, I_K \\
    [0]_K & (I_K+\Delta t\, A_\mu) 
    \end{pmatrix}
    \]
    For unconditional stability in time, it is required for the eigenvalues
    of $I_K+\Delta t A_\mu$ to stay in the unit disk of the complex plane.
    
    More generally, it is possible to use any other time advance scheme, according to the expected order of accuracy or stability domain.
    \subsection{Accuracy criteria and similarity distances between ROM and FOM solutions}
    
In order to quantify the error induced by approximations, we introduce 3 accuracy criteria. 
The first accuracy criterion  is the relative information content (RIC), defined by 
\[
\text{RIC}(K) = \frac{\displaystyle{\sum_{k=K+1}^{\min(d,N)} \sigma_k^2}}{\displaystyle{\sum_{k=1}^{\min(d,N)}\sigma_k^2}},
\]  
quantifies the relative amount of neglected information when truncating the number of modes at rank $K$. 
The truncation rank is determined such that the RIC is inferior to the accuracy threshold~$\varepsilon$. The accuracy threshold $\varepsilon$ is fixed to $10^{-6}$.

The second accuracy criterion  is the relative time residual $\mathcal{R}$. It quantifies the relative error induced by the minimization of the least square cost function (\ref{eq:flo17}) using $A_{\mu}$. It is given by
\[
\mathcal{R}(j)=\frac{\Vert A_{\mu} \mathbb{X}_j-\mathbb{Y}_j \Vert_1}{\Vert  \mathbb{Y}_j \Vert_1}
\]
where $\mathbb{X}_j$ represents the $j^{th}$ column of $\mathbb{X}$ and $\mathbb{Y}_j$ the $j^{th}$ column of $\mathbb{Y}$. The index $j$ is thus linked to the snapshots ($j\in \{1, ..., N\}$). To better draw a parallel between the evolution of this parameter and the capsule dynamics, this parameter will be represented as a function of the non-dimensional time $Vt/\ell$ hereafter.

The third accuracy criteria $\varepsilon_{\text{Shape}}(Vt/\ell)$ measures the difference between the 3D reference capsule shape given by the FOM ($\mathcal{S}_{\text{FOM}}$) and the 3D shape predicted by the ROM ($\mathcal{S}_{\text{ROM}}$). It is defined  at a given non-dimensional time $Vt/\ell$ as the ratio between the modified Hausdorff distance (MHD)  computed between $\mathcal{S}_{FOM}$ and $\mathcal{S}_{ROM}$ and non-dimensionalized by $\ell$
\[
\varepsilon_{\text{Shape}}(Vt/\ell) =\frac{\text{MHD}(\mathcal{S}_{\text{FOM}}(Vt/\ell), \mathcal{S}_{\text{ROM}}(Vt/\ell))}{\ell}
\]
The modified Hausdorff distance is the maximum value of the mean distance between $\mathcal{S}_{\text{FOM}}$ and $\mathcal{S}_{\text{ROM}} $ and the mean distance between $\mathcal{S}_{\text{ROM}}$ and $\mathcal{S}_{\text{FOM}}$ \citep{Dubuisson1994}.

\section{Numerical experimentation on a given configuration}
\label{sec:1example}

\begin{figure}
\centering

(a)\includegraphics[width=0.18\textwidth]{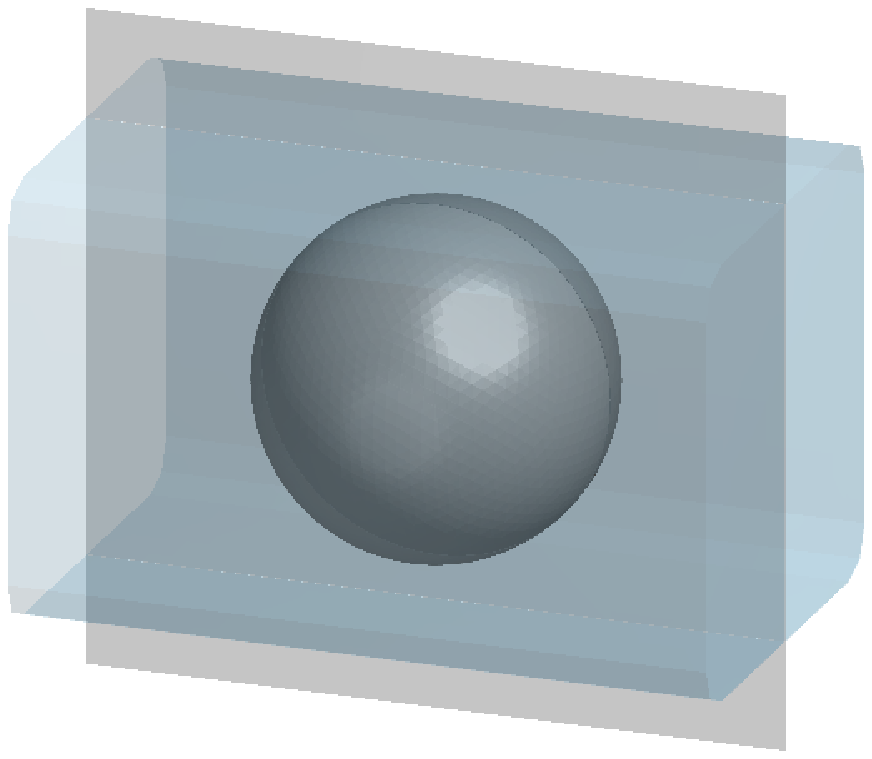}
(b)\includegraphics[width=0.7\textwidth]{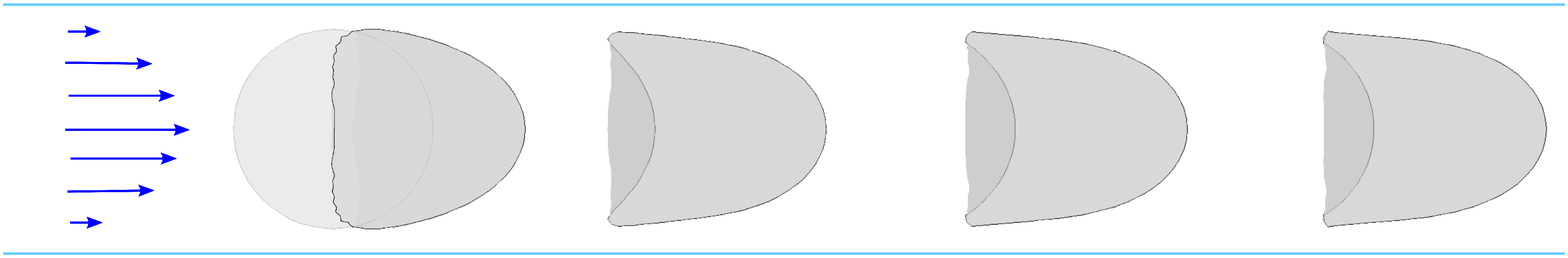}
\caption{Dynamics of a microcapsule flowing in a microchannel with a square cross-section predicted by FOM in the vertical cutting plane represented in grey in (a). The in-plane capsule profiles are shown for $Ca=0.17$ and $a/\ell=0.8$ at the non-dimensional times $Vt/\ell=$ 0, 0.4, 2, 4, 6 in (b). The horizontal lines on (b) represent the channel borders. The capsule will always be represented flowing from left to right.} 
\label{DynamicFOM}
\end{figure}

The method is first applied to a given configuration, in order to set the model parameters  and to study its stability and precision. We consider the dynamics of an initially spherical capsule flowing in a microchannel when $Ca=0.17$ and $a/\ell=0.8$.  The time step between each snapshot $\Delta t$ equals to 0.04. The dynamics predicted by the FOM is illustrated  in Fig. \ref{DynamicFOM} up to a nondimensional time $VT/\ell=10$. As the capsule flows, its membrane is gradually deformed by the hydrodynamic forces inside the channel during a temporary time until a steady state is reached. We assume that the capsule has reached its steady-state shape, when the surface area of the capsule varies by less than $5\times10^{-4}\times(4 \pi a^2)$ over a non-dimensional time $Vt/\ell=1$. For $(Ca=0.17$, $a/\ell=0.8)$, the steady state is reached at $Vt/\ell=2.6$ and is characterized by a parachute capsule shape (Figure \ref{DynamicFOM}). 

\subsection{Proper orthogonal decomposition, truncation and modes}

\begin{figure}
\centering
\includegraphics[width=0.55\textwidth]{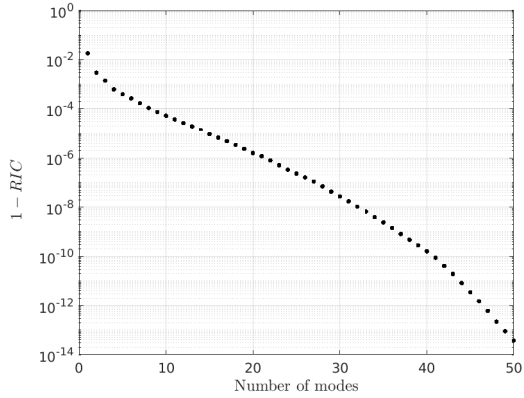}
\caption{Evolution of the relative amount of neglected information 1-RIC, as a function of the number of modes considered ($Ca=0.17, a/\ell=0.8$).}
\label{RIC}
\end{figure}

The singular value decomposition is first applied to the displacement snapshot matrix. To determine the truncation rank, the evolution of 1- RIC is illustrated in Figure~\ref{RIC} as a function of number of modes considered. The RIC is about 1\% only with one mode. The more modes is kept, the less information is neglected.  In the following, we fix the number of modes to 20. The accuracy threshold $\varepsilon$ is thus equal to $10^{-6}$.

\begin{figure}
\centering
(a)\includegraphics[width=0.28\textwidth]{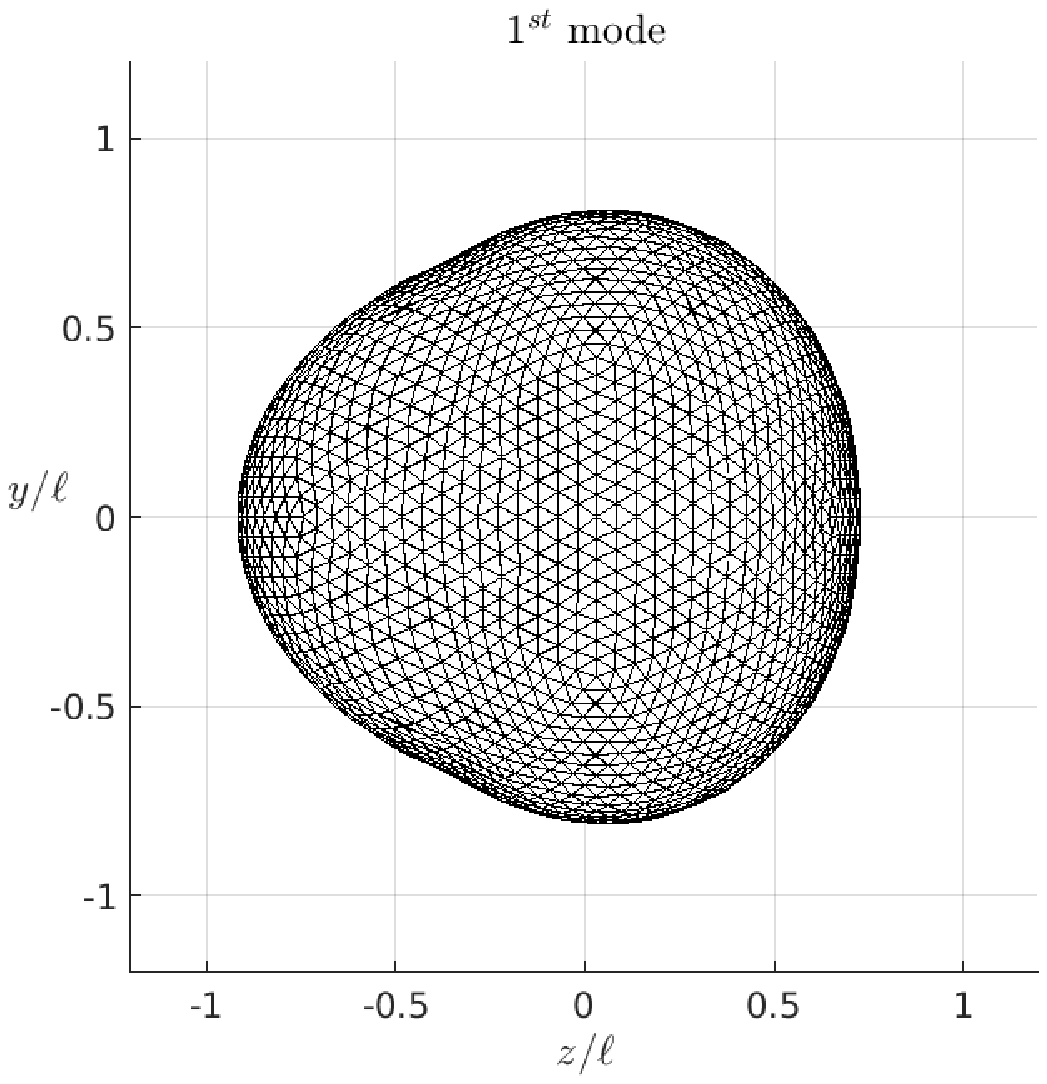}
(b)\includegraphics[width=0.28\textwidth]{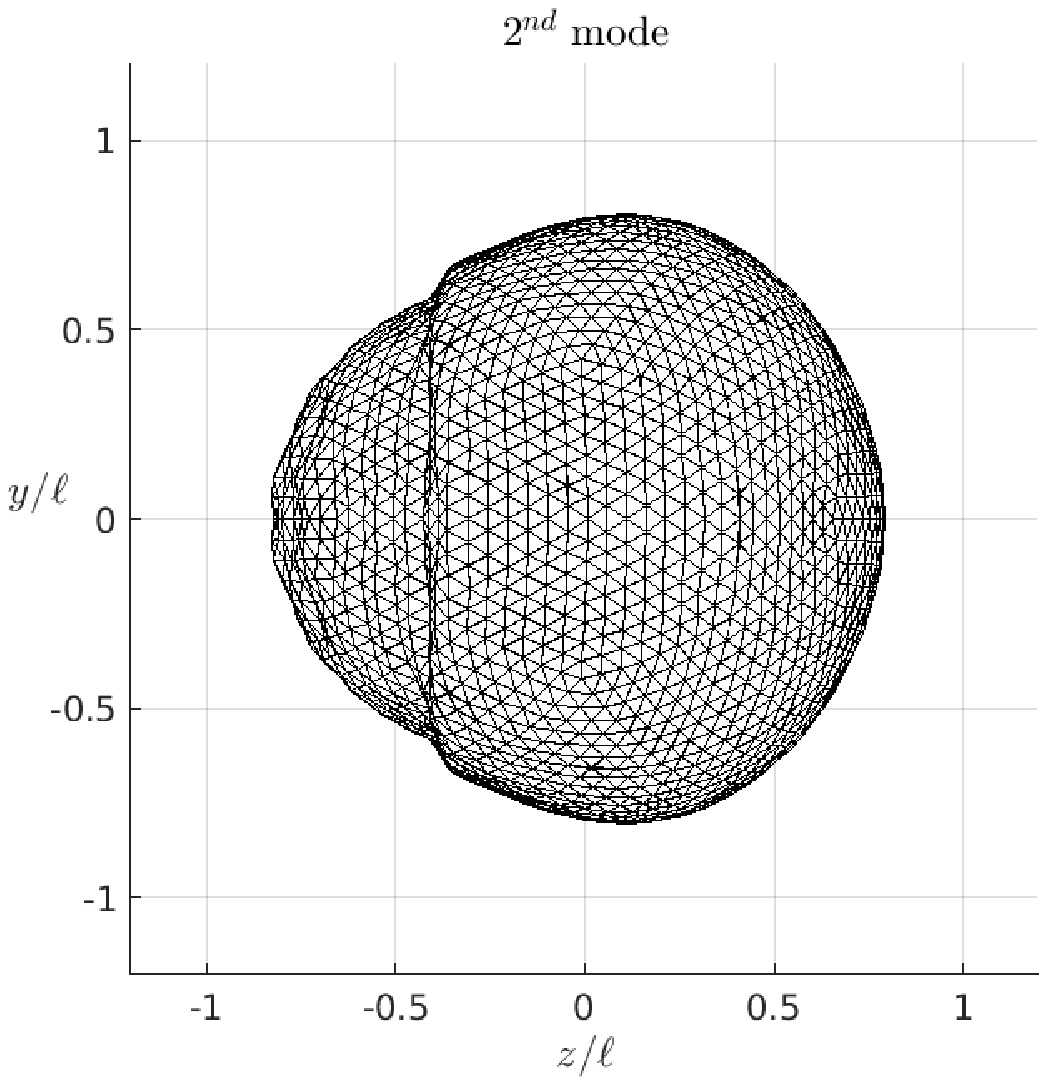}
(c)\includegraphics[width=0.28\textwidth]{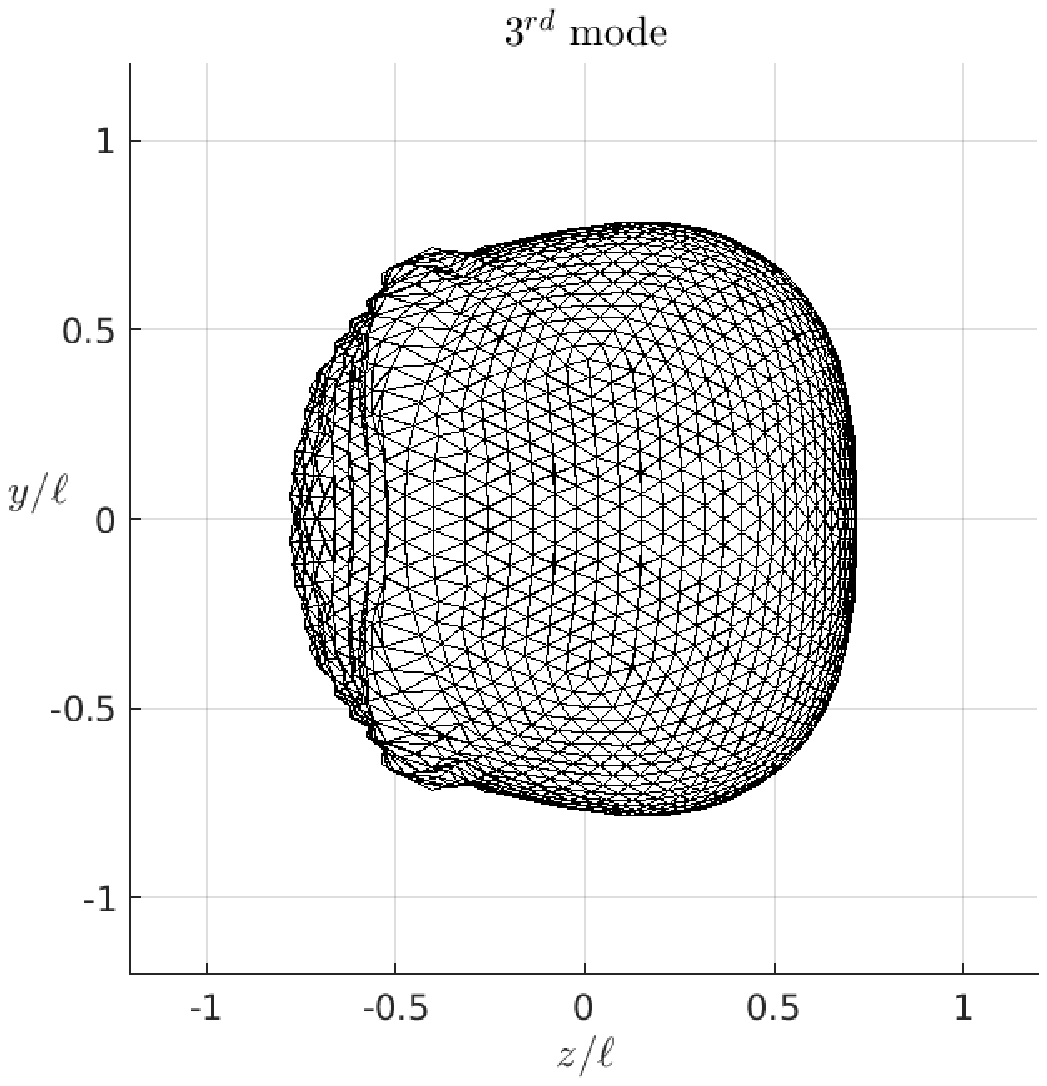}
(d)\includegraphics[width=0.28\textwidth]{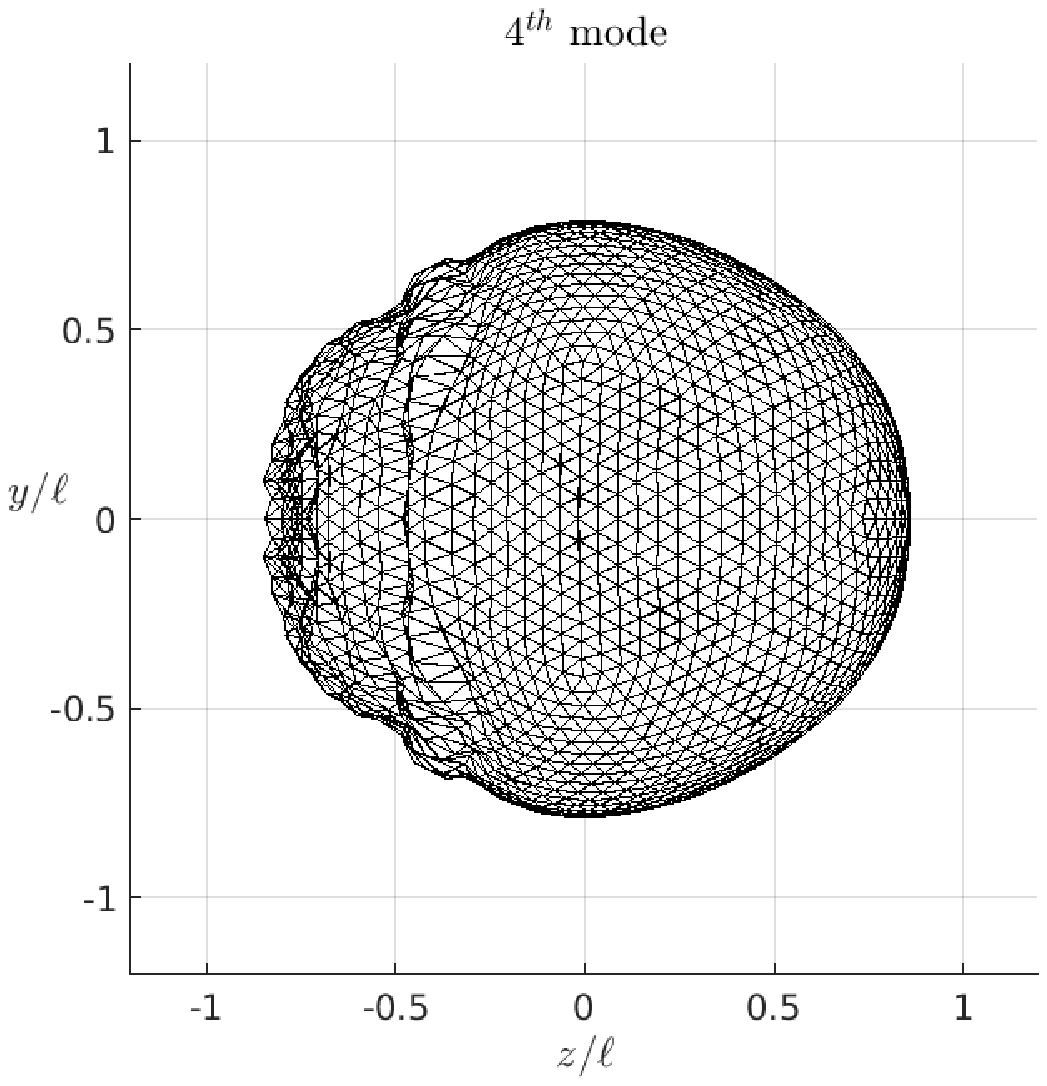}
(e)\includegraphics[width=0.28\textwidth]{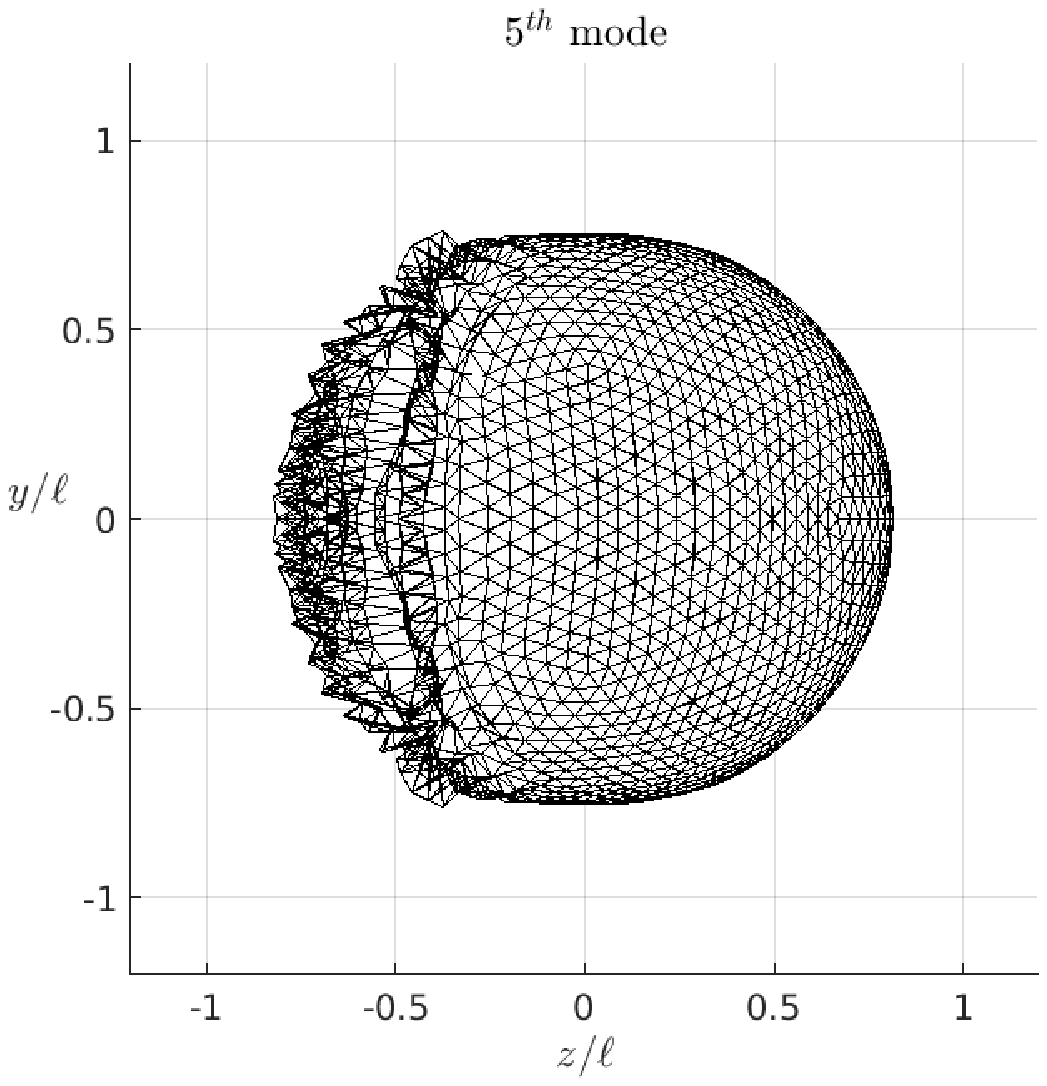}
(f)\includegraphics[width=0.28\textwidth]{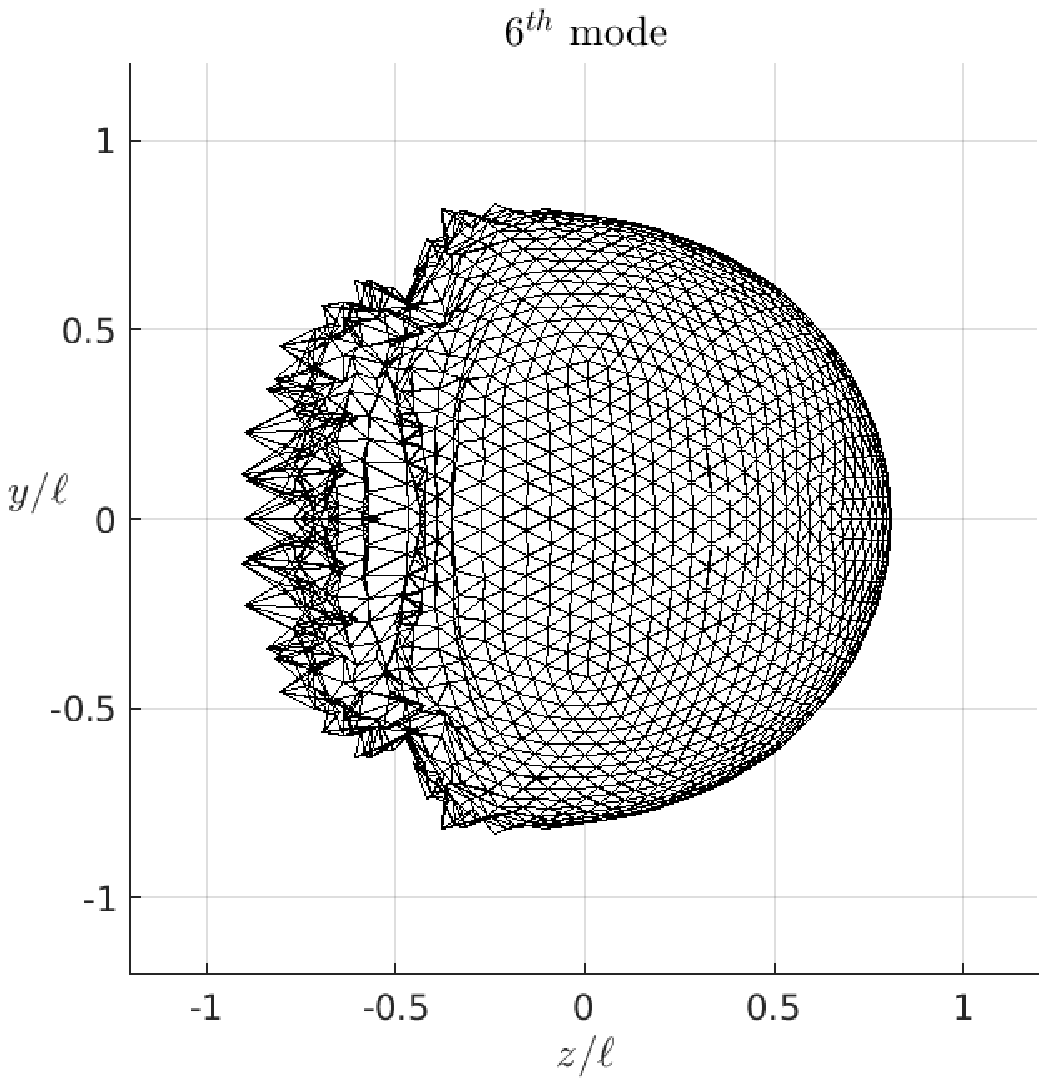}
\caption{Representation of the first six modes of the capsule dynamics when $a/\ell =0.80$ and $Ca=0.17$. To be vizualized the modes of displacement were added to the sphere of radius 1 and amplified by a factor 2.} 
\label{modesRepresentation}
\end{figure}

The modes are determined from the displacement snapshot matrix. They are added to the sphere of radius 1 and amplified by a factor 2 to be  visualized. The first six modes are represented in Figure~\ref{modesRepresentation} for ($Ca=0.17, a/\ell=0.8$).
The first six modes are mostly dedicated to change the shape of the rear of the capsule. The following modes appear to become noisy (not shown). However, these modes are not negligible, if one wants to get an accuracy of $10^{-6}$.

\subsection{Dynamic Mode Decomposition: empirical regularization}

\begin{figure}
\centering
\includegraphics[width=0.55\textwidth]{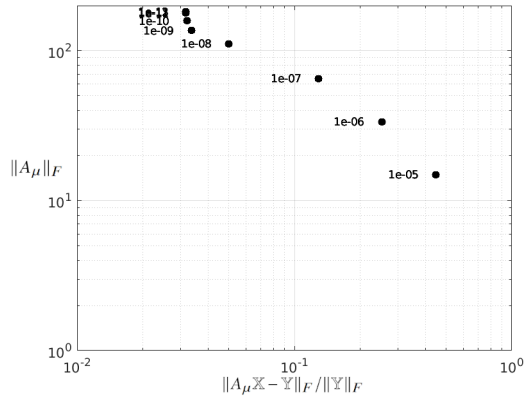}
\caption{Evolution of the norm solution $\Vert A_{\mu}\Vert_F$ as a function of
the least square error $\Vert A_{\mu}\mathbb{X}-\mathbb{Y}\Vert_F / \Vert \mathbb{Y}\Vert_F$
 when the number of modes is fixed to 20 and $(Ca=0.17, a/\ell=0.8)$.} 
\label{muDetermination}
\end{figure}

Before determining the matrix $A$, we check the condition number of the matrices $\mathbb{X}$ and $\mathbb{XX^T}$. They are respectively equal to $7.9\times 10^4$ and $6.2 \times 10^9$. The condition numbers of these matrices are very high and the matrix $A$, determined by solving (\ref{eq:flo20}), may be sensitive to perturbations or noise. To improve the robustness, a Tikhonov regularization is applied to solve the least-square problem (\ref{eq:flo17}) and the matrix $A_{\mu}$ is computed using (\ref{eq:flo22}), which depends on the regularization coefficient $\mu$.
To determine the optimal value of $\mu$, the relative least square error $\Vert A_{\mu}\mathbb{X}-\mathbb{Y}\Vert_F/\Vert\mathbb{Y}\Vert_F$ is represented according to the norm solution $\Vert A_{\mu}\Vert_F$ when 20 modes are considered and when $\mu$ is varied between \review{$10^{-12}$ and $10^{-5}$} (Figure \ref{muDetermination}). 
The least square error $\Vert A_{\mu}\mathbb{X}-\mathbb{Y}\Vert_F$ and the norm solution $\Vert A_{\mu}\Vert_F$ are minimal when $\mu=10^{-9}$. In the following, $\mu$ is thus fixed to $\mu=10^{-9}$ and the number of modes to 20. 

\subsection{Validity check of the ROM: spectral study of the resulting matrix}

\begin{figure}
\centering
\includegraphics[width=0.55\textwidth]{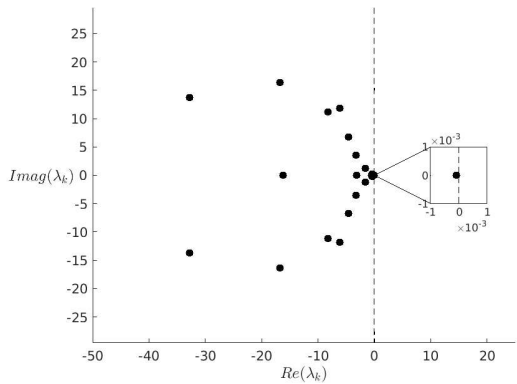}
\caption{Eigenvalues $\lambda_k$ of $A_{\mu}$ when 20 modes are considered, $\mu=10^{-9}$ and $(Ca=0.17, a/\ell=0.8)$. } 
\label{eigenvalues}
\end{figure}

In order to detect anomalies, a spectral analysis of the reduced-order model learned by the DMD method is carried out. The spectrum of the matrix $A_{\mu}$ is represented in Figure~\ref{eigenvalues}. All the eigenvalues $\lambda_k$ of the matrix $A_{\mu}$ have non-positive real parts. The resulting linear ROM is thus stable. 

\begin{figure}
\centering
\includegraphics[width=0.55\textwidth]{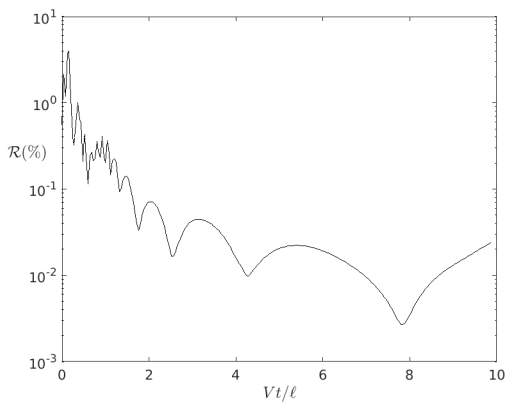}
\caption{Temporal evolution of  the normalized time residual when 20 modes are considered, $\mu=10^{-9}$ and $(Ca=0.17, a/\ell=0.8)$.} 
\label{ErrorModel}
\end{figure}

The temporal evolution of the residual $\mathcal{R}$ (Figure~\ref{ErrorModel}) shows that the error is less than 4\%. The maximal value is reached
at the beginning of the simulation ($Vt/\ell < 0.3$) and $\mathcal{R}$ decreases afterwards. 
When $Vt/\ell \lesssim 2$, i.e. before the capsule has reached its steady state, high  frequency oscillations are observed. This probably means that a high frequency mode is neglected, even if 20 modes are considered. For $Vt/\ell >6$, $\varepsilon_{ROM}$ is of order $0.01\%$. The stationary state is thus well predicted by the model and the error during the transient stage is more than acceptable. 

\subsection{ROM online stage and accuracy assessment}

\begin{figure}
\centering
\includegraphics[width=0.9\textwidth]{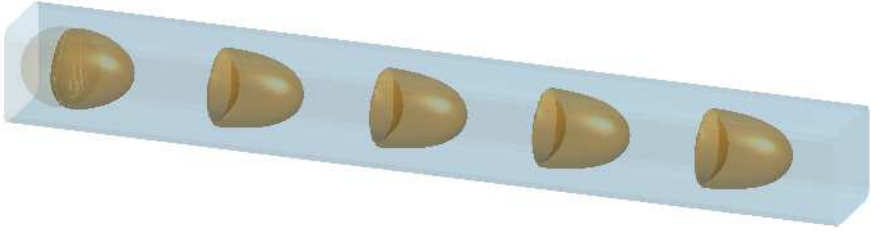}
\caption{Dynamics of a microcapsule flowing in microchannel with a square cross-section predicted by ROM at the non-dimensional time $Vt/\ell=$ 0.4, 2.8, 5.2, 7.6, 10 when $Ca=0.17$ and $a/\ell=0.8$. The initial spherical capsule is shown on the left by transparency. The number of modes is fixed to 20 and $\mu=10^{-9}$.} 
\label{DynamicsROM3D}
\end{figure}

\begin{figure}
\centering
\includegraphics[width=\textwidth]{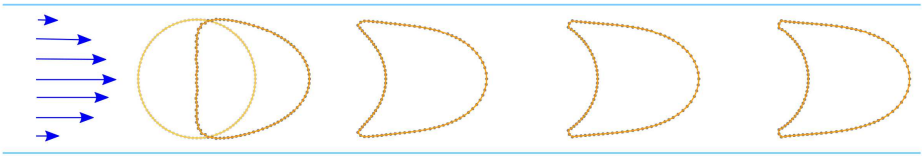}
\caption{Comparison of the capsule contours given by the FOM (dotted line) and estimated by the ROM (orange line). The capsule is shown for $(Ca=0.17, a/\ell=0.8$ at the non-dimensional time $Vt/\ell=$ 0, 0.4, 2, 4, 6. The horizontal lines represent the channel borders. The number of modes is fixed to 20 and $\mu=10^{-9}$.} 
\label{CompProfils}
\end{figure}



\begin{figure}
\centering
a)\includegraphics[width=0.6\textwidth]{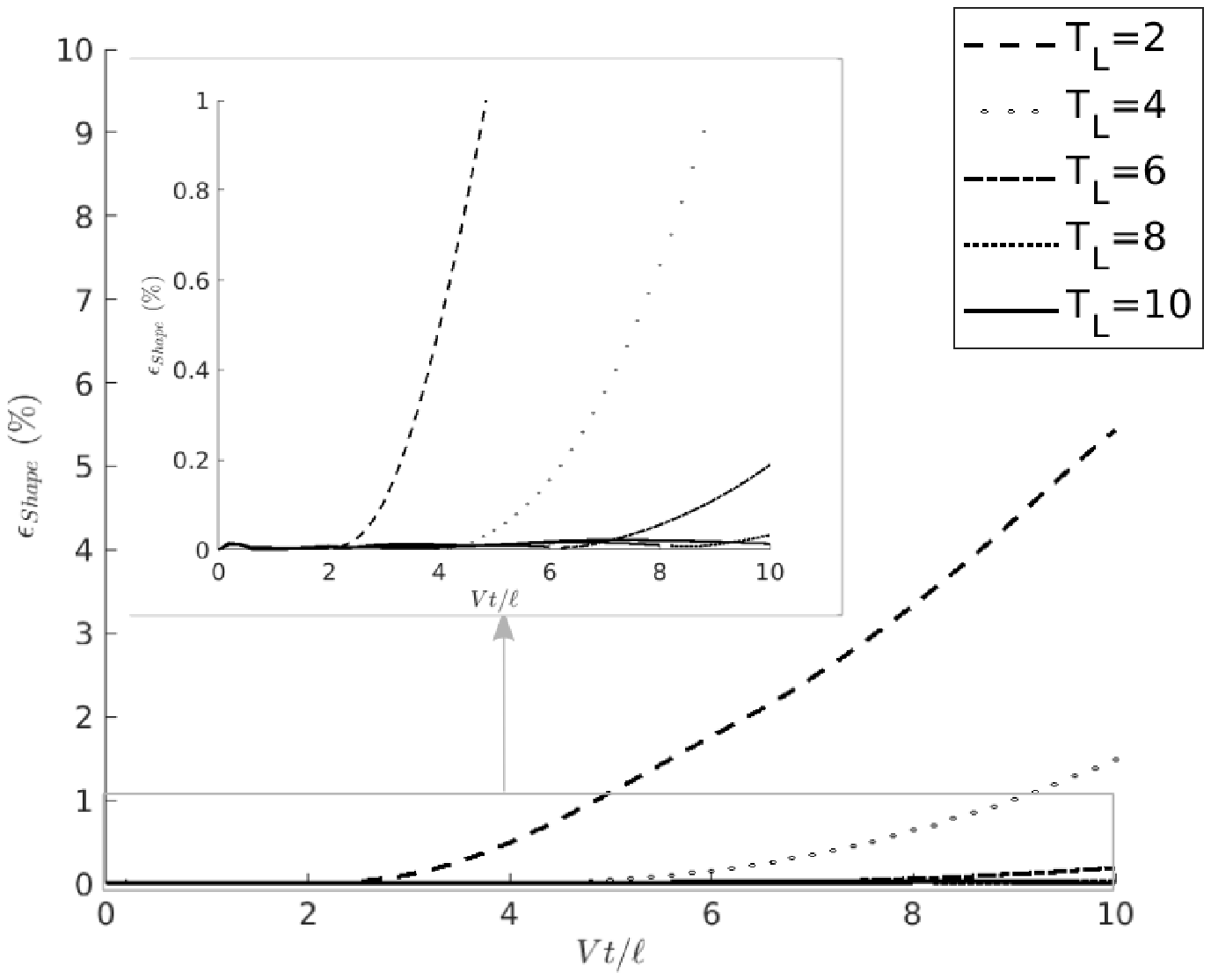}\\
b)\includegraphics[width=0.6\textwidth]{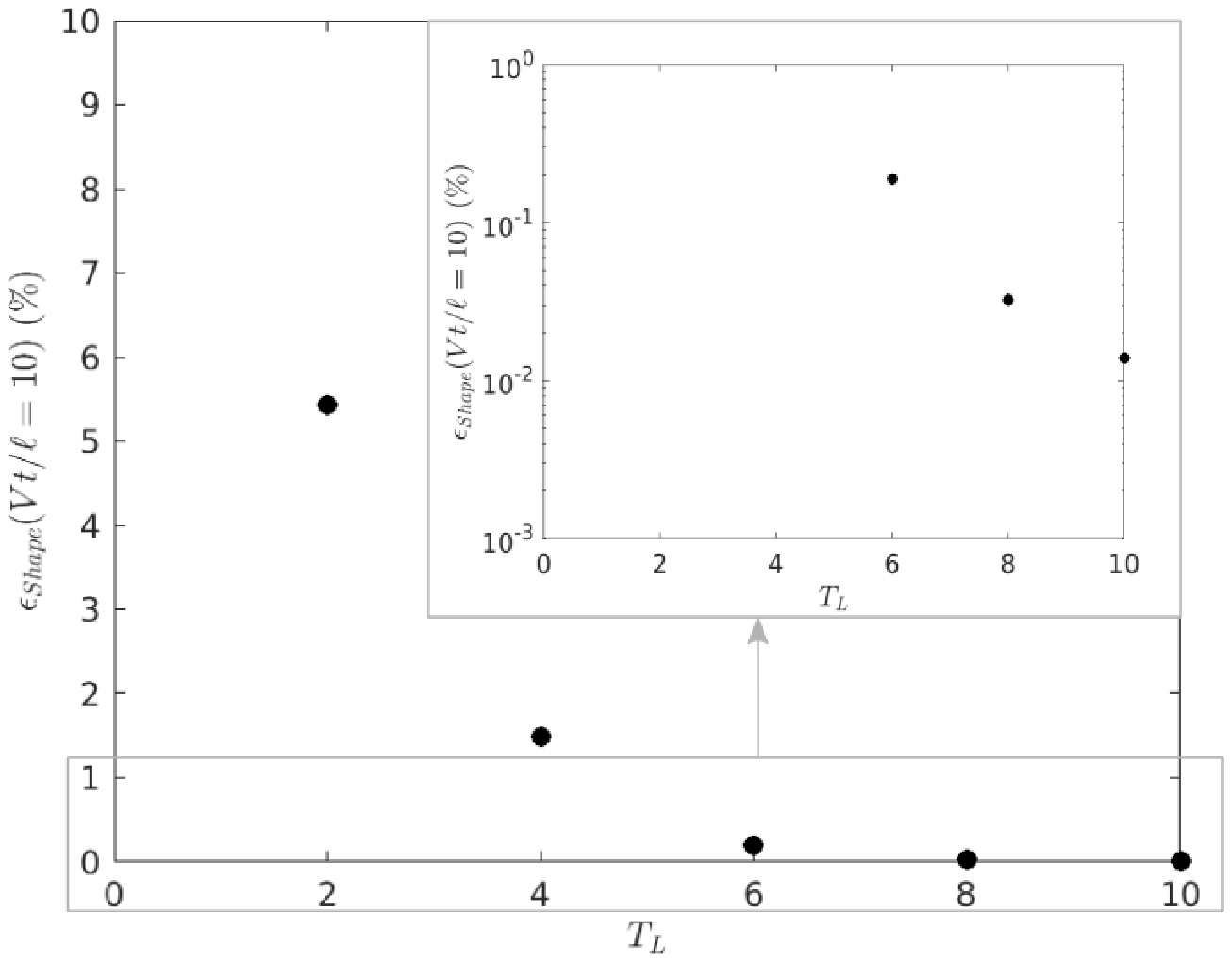}\\
c)\includegraphics[width=0.95\textwidth]{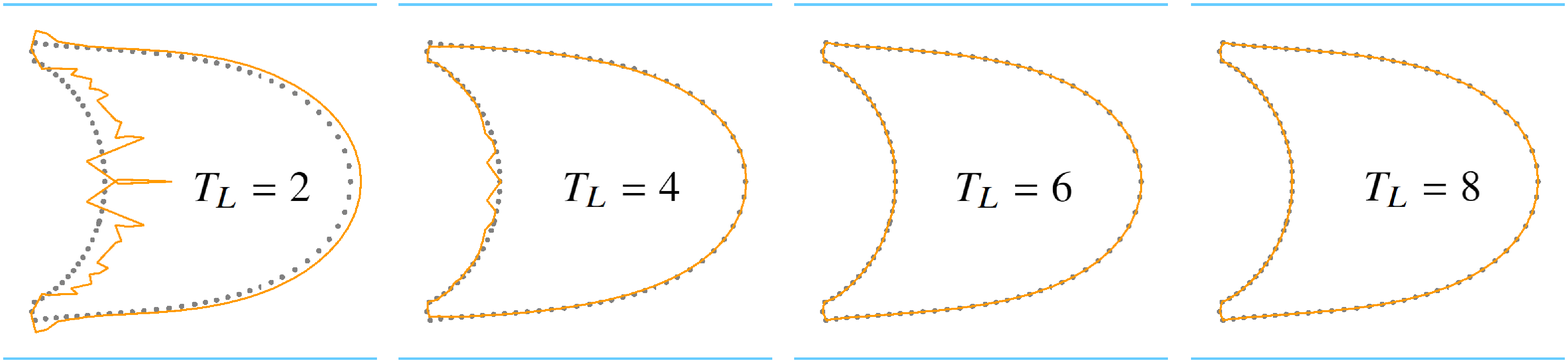}
\caption{\review{ a) Influence of the learning time $T_L$ on the temporal evolution of the error on the capsule shape $\varepsilon_{\text{Shape}}$. The error during the learning time is shown in solid line. b) Evolution of $\varepsilon_{\text{Shape}}$ measured at $Vt/\ell=10$ as a function of the learning time $T_L$. c) Comparison of the capsule contours given by the FOM (dotted line) and estimated by the ROM (orange line) for the different learning times $T_L$. For this case, the parameters are 20 modes, $\mu=10^{-9}$, $Ca = 0.17$ and $a/\ell=0.8$. } } 
\label{TimeLearning}
\end{figure}

The displacement of all the nodes of the capsule mesh estimated by the ROM is then added to the corresponding node of the sphere of radius 1 to visualize the temporal evolution of the capsule shape in three dimensions. Figure \ref{DynamicsROM3D} shows the capsule dynamics for the reference case ($Ca=0.17, a/\ell=0.8$). The ROM allows us to reproduce the capsule deformation from the initial state up to the parachute-shaped steady state. 
For the FOM and the ROM, the capsule profile is then determined in the
 cutting plane passing through the middle of the microchannel. Figure \ref{CompProfils} shows that the two capsule profiles perfectly overlap  at $Vt/\ell=0, 0.4, 2, 4, 6$.
The temporal evolution of $\varepsilon_{\text{Shape}}$ is shown in Figure \ref{TimeLearning}a. The maximum value of the error committed on the 3D shape $\varepsilon_{\text{Shape}}$ equals to 0.022\%.
 The error on the capsule shape $\varepsilon_{\text{Shape}}$ is thus negligible. The deformation of the capsule from its initially spherical shape to its steady state over an non-dimensional time $Vt/\ell=10$ can thus be estimated very precisely with the developed reduced-order model.

\review{\subsection{Sensitivity of the ROM on the learning time  }

The DMD method predicts the capsule displacement at time $t^{n+1}$ from that at time $t^n$. The model has been constructed until now by considering the dynamics of the capsule over a non-dimensional time $Vt/\ell$ of 10.

In order to study its influence, we increase the learning  $T_L$, i.e. the non-dimensional time over which the model is trained, from 2 to 8 and estimate the capsule dynamics using the ROM up to a non-dimensional time $Vt/\ell$ of 10. The number of modes is always equal to 20 and $\mu=10^{-9}$. The estimated shape is then compared to the one simulated with the FOM. The time evolution of $\varepsilon_{\text{Shape}}$ is shown in Figure \ref{TimeLearning}a. The error between the 3D shape increases, as soon as we exceed the learning time. The longer the learning time, the smaller the error $\varepsilon_{\text{Shape}}$ at $Vt/\ell=10$ (Figure \ref{TimeLearning}b). The comparison of the capsule profile estimated by the ROM at $Vt/\ell=10$ with the one simulated with the FOM (Figure \ref{TimeLearning}c) confirms that considering only the dynamics of the capsule up to a non-dimensional time $Vt/\ell$ of 2 is not sufficient. In fact, the capsule is still in the transient phase at $Vt/\ell=2$. When $T_L=4$, a small difference persists in the parachute at $Vt/\ell=10$ and $\varepsilon_{\text{Shape}}=1.5\%$. However, when the learning time is increased above 6, $\varepsilon_{\text{Shape}}$ falls below 0.19\% at $Vt/\ell=10$.  This learning time is sufficient to estimate the overall shape of the capsule.}

\section{\review{Space-time ROM accuracy assessment over the full parameter sample set}}
\label{sec:database}
\begin{figure} 
\centering
\includegraphics[width=0.55\textwidth]{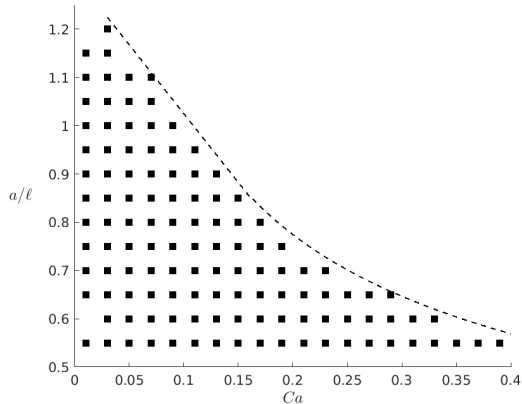}
\caption{Values of $Ca$ and $a/\ell$ included in the training database. The dotted line delimits the domain where a steady-state capsule deformation exists for capsules following the neo-Hookean law.}
\label{DB}
\end{figure}

The capillary number  $Ca$ and the aspect ratio $a/\ell$ are now considered as variable parameters.
A database of 119 simulations of the deformation of an initially spherical capsule in a microchannel has been generated using the FOM with the same time step and mesh size as in section \ref{sec:1example}. Figure~\ref{DB} shows the different values of $Ca$ and $a/\ell$ for which the simulations have been computed to create the training database. 
When the capsule initial radius is close to or larger than the microchannel cross-dimension ($a/\ell \geq 0.90$), the capsule is predeformed into a prolate spheroid to fit in the channel. For a given $a/\ell$, a limit value of $Ca$ exists beyond which a capsule does not reach a steady-state (Figure~\ref{DB}). This is due to the softening behavior of the neo-Hookean law. 

\begin{figure}
\centering
(a)\includegraphics[width=0.45\textwidth]{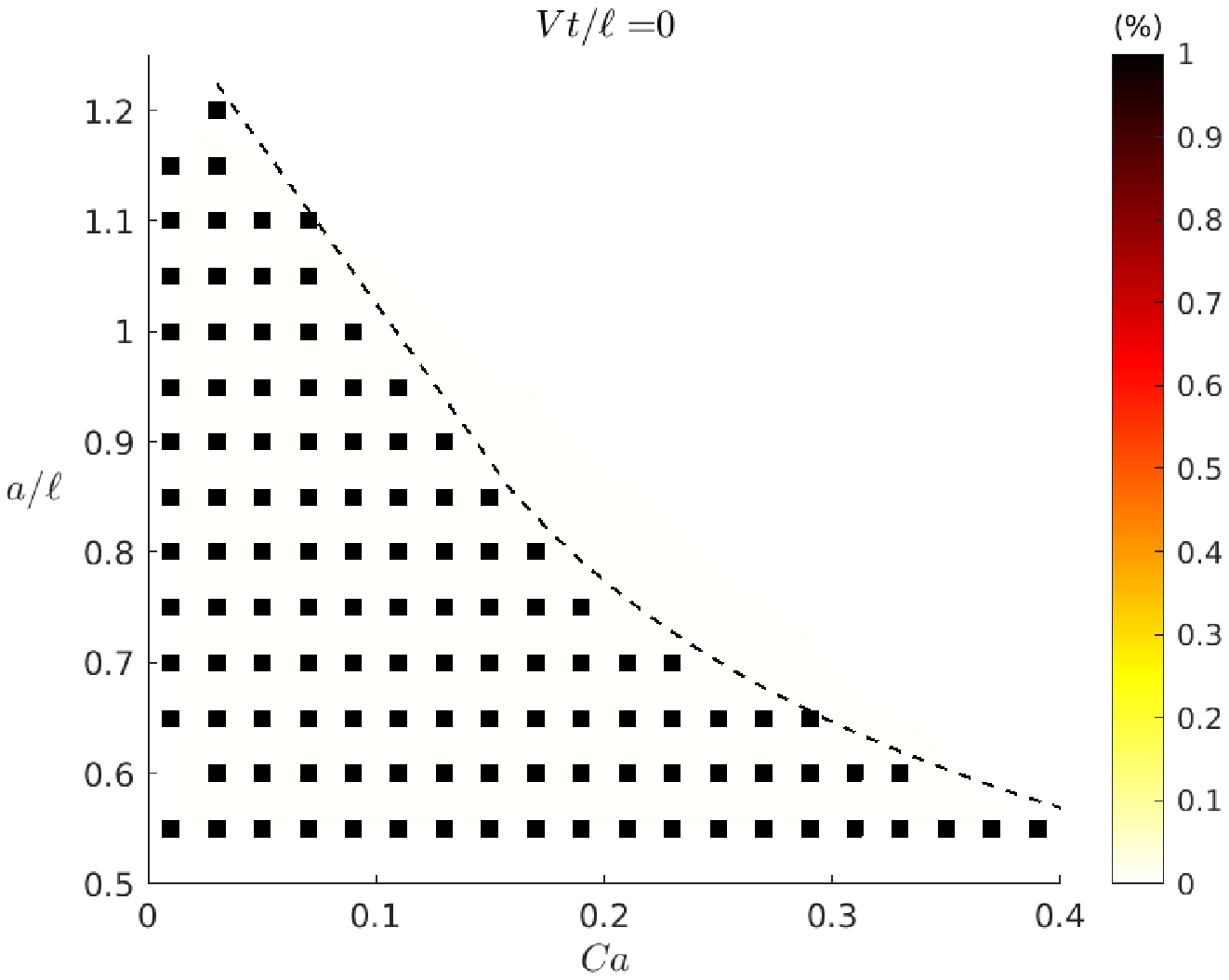}
(b)\includegraphics[width=0.45\textwidth]{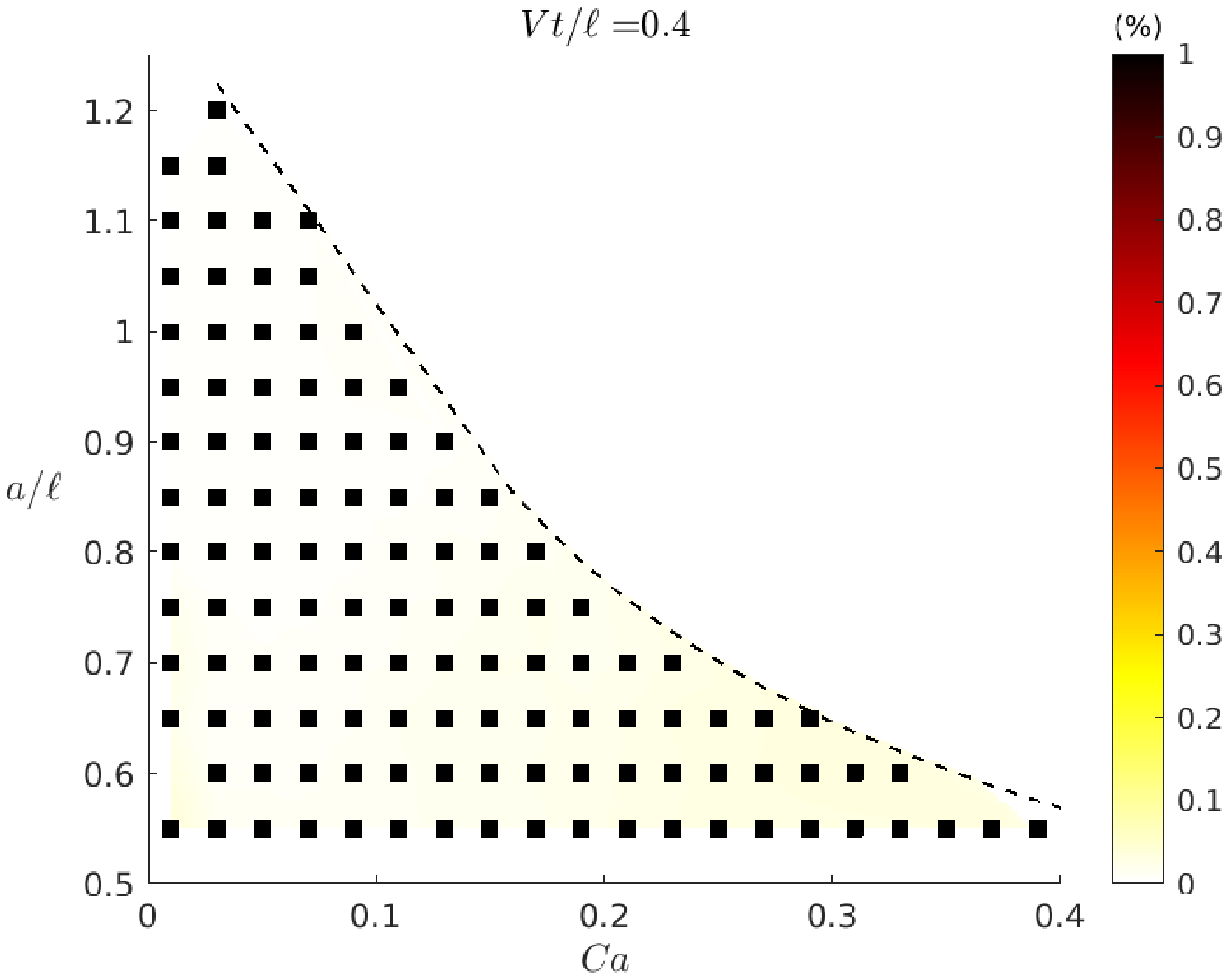}
(c)\includegraphics[width=0.45\textwidth]{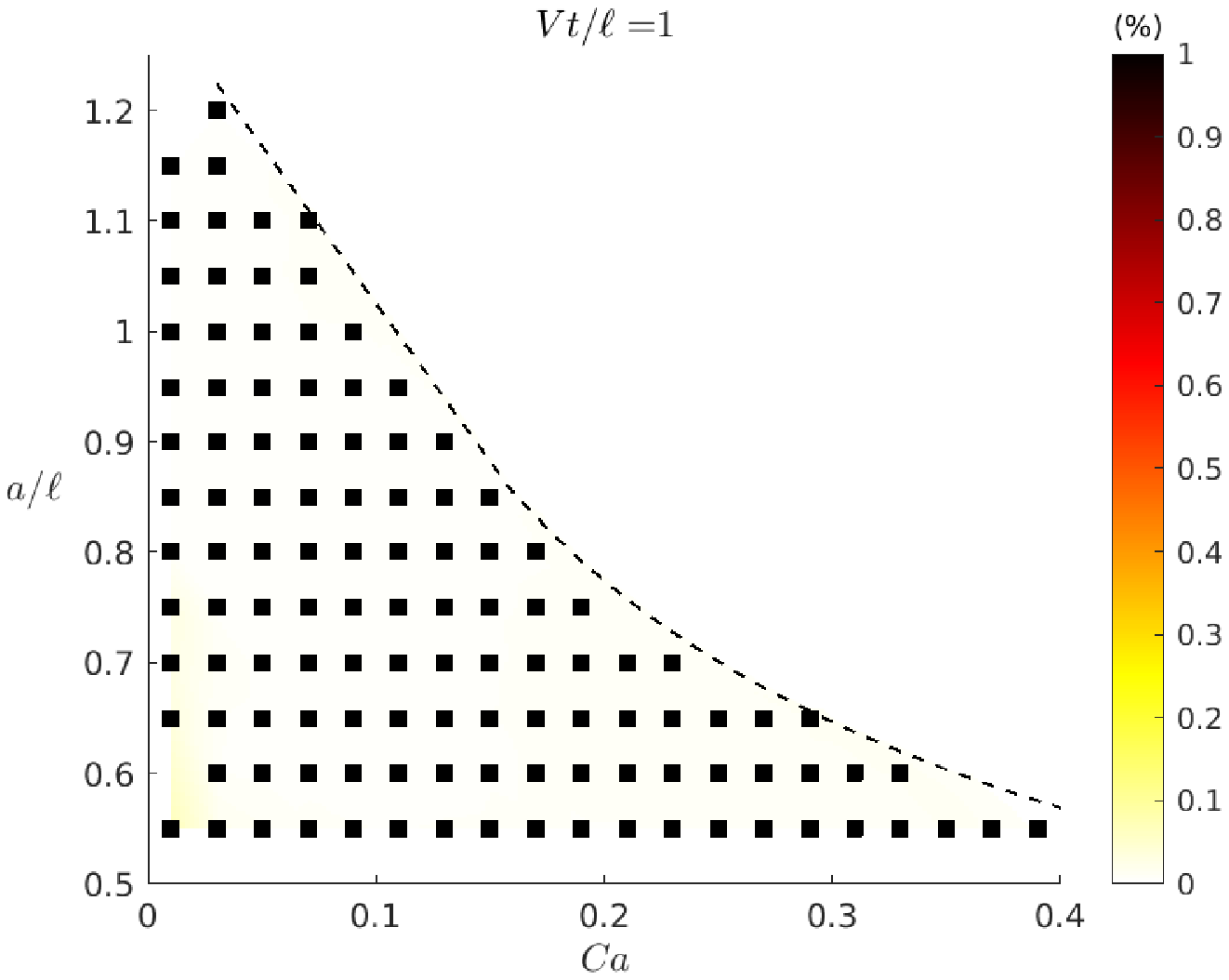}
(d)\includegraphics[width=0.45\textwidth]{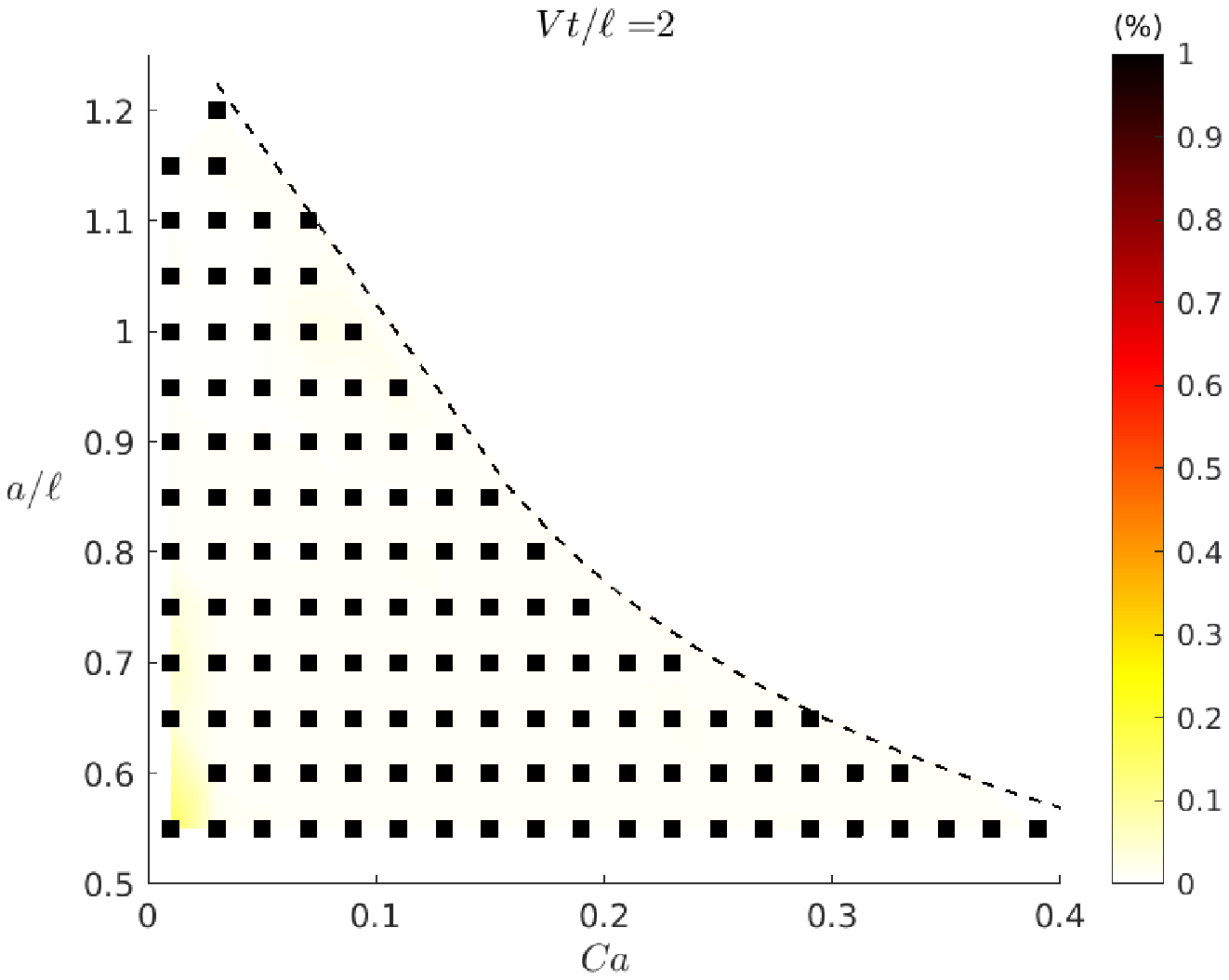}
(e)\includegraphics[width=0.45\textwidth]{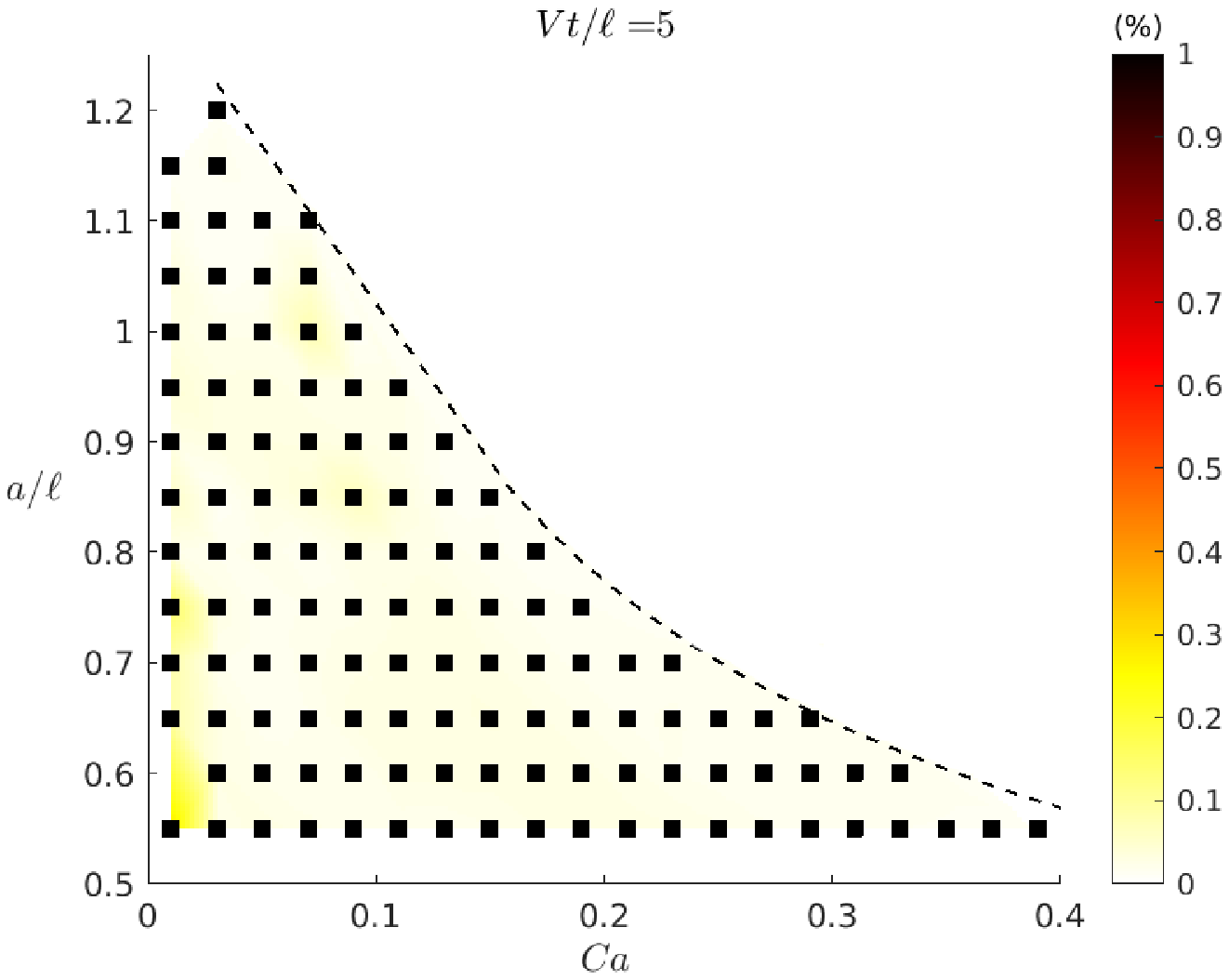}
(f)\includegraphics[width=0.45\textwidth]{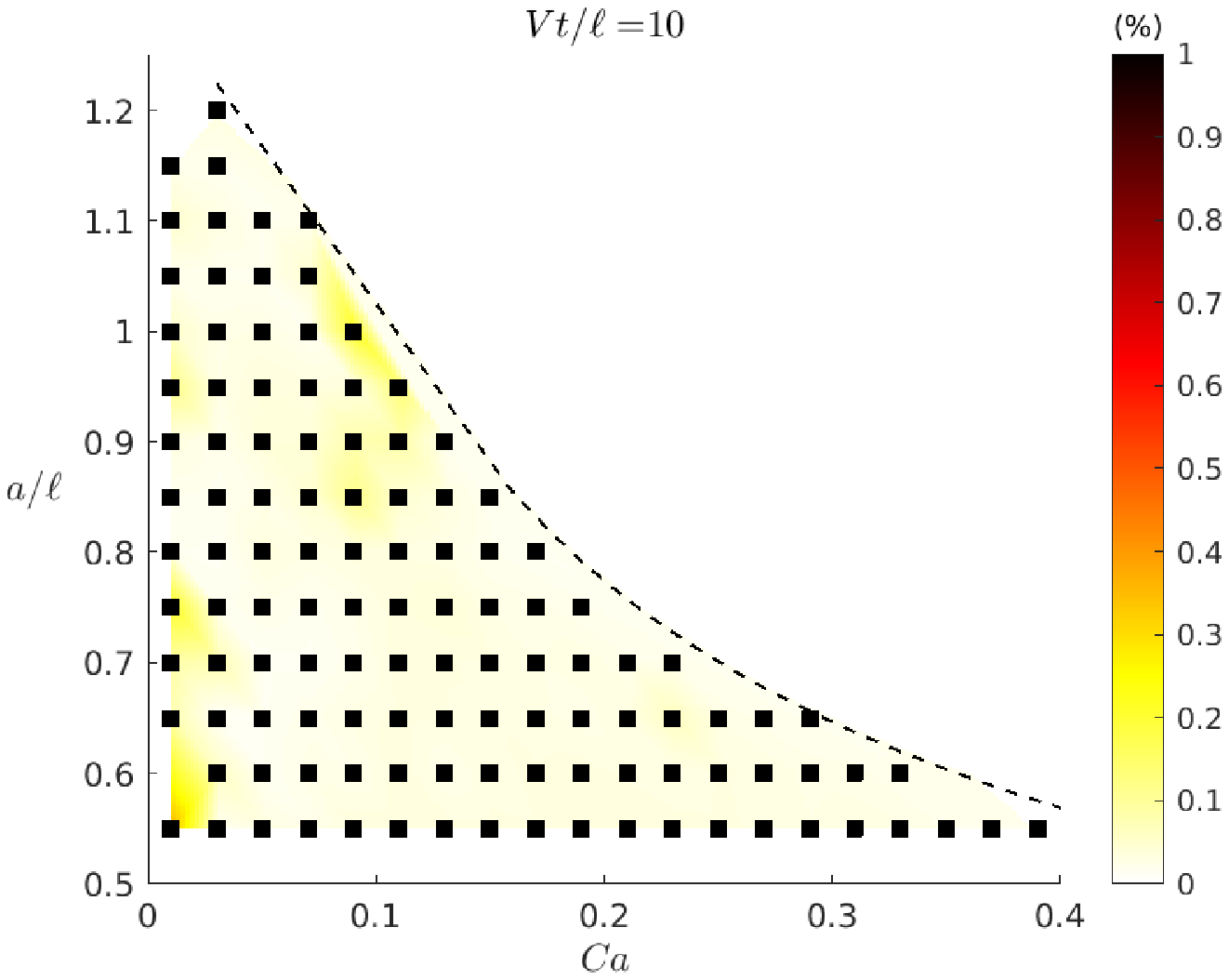}
\caption{Heat maps of $\epsilon_{Shape}$ on the training database as a function of $Ca$ and $a/\ell$ at (a) $\dot{\gamma}t$=0, (b) 0.4, (c) 1, (d) 2, (e) 5, (f) 10. 20 modes and $\mu=10^{-9}$ are considered. The dotted line delimits the domain where a steady-state capsule deformation exists.} 
\label{ErreurBaseA20}
\end{figure}

For all the couples ($Ca, a/\ell$) of the training database, the capsule shape is reconstructed from the ROM results at given non-dimensional times and compared to the shape predicted by the FOM at the same non-dimensional time. The evolution of the error committed on the capsule shape $\varepsilon_{Shape}$ on the full database is illustrated in Figure~\ref{ErreurBaseA20} at $Vt/\ell=0, 0.4, 1, 2, 5, 10$.  $\varepsilon_{Shape}$ is null at $Vt/\ell=0$. The ROM is therefore able to  predict the initial capsule shape correctly, whether it is spherical or slightly ellipsoidal. Until $Vt/\ell\leq2$, $\varepsilon_{Shape}$  essentially remains zero on the majority of the database. Otherwise, it is equal to 0.15\% at maximum. At $Vt/\ell = 5$ and 10, the error $\varepsilon_{Shape}$ slightly increases for most of the couples ($Ca, a/\ell$) of the database. It remains fully acceptable since it is equal to 0.35\% at maximum. When considering 20 modes and $\mu = 10^{-9}$, the developed ROM allows us to estimate with great precision the dynamics of an initially spherical capsule in a  microchannel with a square cross-section. 

\begin{figure}
\centering
\includegraphics[width=0.55\textwidth]{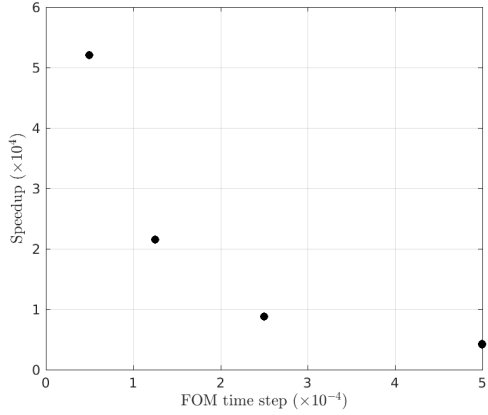}
\caption{Evolution of the speedup as function to the time step imposed to simulate the capsule dynamics with the FOM ($a/\ell=0.7$).}
\label{Speedup} 
\end{figure}

To respect the stability condition (see Equation~\ref{StabilityCondition}), the time step imposed to simulate the capsule dynamics with the FOM decreases, when the $Ca$ decreases. The lower the $Ca$, the longer the simulation lasts (Figure~\ref{SimuTime}). The time needed to calculate the capsule shape and write the results was estimated on the same workstation used to simulate and generate the result files with the FOM (2-CPU Intel\textsuperscript{\textregistered} Xeon\textsuperscript{\textregistered} Gold 6130, 2.1 GHz). The speedup is the ratio between the FOM runtime and the ROM runtime. Its evolution according to the FOM time step is illustrated in Figure~\ref{Speedup}. It was estimated from the ROM and FOM simulation time obtained when $a/\ell=0.7$. The speedup varies between 52106 for a FOM time step of $10^{-4}$ (i.e for the lowest value of $Ca$ tested) and 4200 for $5 \times 10^{-4}$ (i.e $Ca\leq 0.05$). It is thus possible to estimate the capsule dynamics very precisely with the developed ROM, while considerably reducing  the computational time.

\review{Another significant advantage is the gain in storage of the simulation results. By storing only the reduced variables  $\bm{\alpha}$, $\bm{\beta}$, the modes $\lbrace \phi_k\rbrace $ and the initial position of the nodes of each couple $\theta=(Ca,a/\ell)$, the training database is reduced from 1.9 GB, when computed with the FOM, to 0.15 GB with the ROM. It can therefore be more easily shared. }

\section{\review{Full space-time-parameter ROM (for any admissible  parameter value)}}
\label{sec:interpolation}
\review{
\subsection{General methodology}

It is here again assumed that a training database of $N$ precomputed FOM results is available. Now we would like to derive a ROM for any
parameter couple $\btheta=(Ca,a/\ell)$ in the admissible parameter domain. The proposed space-time-parameter ROM is made of two steps. The first step consists in predicting the space-time solution $\{\bu\}(t;\btheta)$ 
by means of a robust interpolation procedure. The second step consists in deriving a ROM in the form of a low-order dynamical system
by using the predicted solutions of the first step as training data. Then we apply the former procedure detailed in Section \ref{sec:ROM}. Below we give a detailed explanation of the two steps. \medskip

\textbf{Step 1: predictor step.} Considering a parameter couple $\btheta$, we first search the three nearest neighbor parameters in the sample set that form a nondegenerate triangle in the plane $(Ca,a/\ell)$. Let us denote them by $\btheta_1$, $\btheta_2$ and $\btheta_3$. We will define a linear operator in the triangle $(\btheta_1,\btheta_2,\btheta_3)$. For that, let us introduce the barycentric coordinates $(\lambda_1,\lambda_2,\lambda_3)$, $\lambda\in[0,1]$, $i=1,2,3$ such that
\begin{align}
   \lambda_1 + \lambda_2 + \lambda_3 & = 1, \label{eq:baryc1}\\
   \btheta_1 \lambda_1 + \btheta_2 \lambda_2 + \btheta_3\lambda_3 & = \btheta. \label{eq:baryc2}
\end{align}
The $3\times 3$ linear system~\eqref{eq:baryc1},\eqref{eq:baryc2} is invertible as soon as the triangle $(\btheta_1,\btheta_2,\btheta_3)$ is nondegenerate. Notice that the $\lambda_i$ ($i=1,2,3$) are actually functions of $\btheta$.
Let us now denote by $\{u_1\}$, $\{u_2\}$ and
$\{u_3\}$ the displacement fields for the parameter vectors
$\btheta_1$, $\btheta_2$ and~$\btheta_3$ respectively. Then we can consider the predicted velocity field $\hat\bu(t;\btheta)$
defined by
\begin{equation}
\{\hat u\}(t,\btheta) =
\lambda_1 \{u_1\}(t) + \lambda_2 \{u_2\}(t) + \lambda_3 \{u_3\}(t).
\label{eq:hatv}
\end{equation}

\textbf{Step 2: low-order dynamical system ROM.}
Expression~\eqref{eq:hatv} can be evaluated at some discrete instants
in order to generate new training data. Then the SVD-DMD ROM methodology presented in Section~\ref{sec:ROM} can be applied to these data to get a reduced dynamical system in the form
\begin{align*}
    & \dot\balpha(\btheta) = \bbeta(\btheta), \\
    & \dot\bbeta(\btheta) = A_\mu(\btheta)\, \bbeta(\btheta).
\end{align*}

We also have a matrix $Q(\btheta)$ of orthogonal POD modes and we can go back to the high-dimensional physical space by the standard operations
\begin{equation}
\{\hat u\}(t,\btheta) \approx Q(\btheta)\, \balpha(t,\btheta),\quad
\{\hat v\}(t,\btheta) \approx Q(\btheta)\, \bbeta(t,\btheta).
\end{equation}
Notice that the capsule position field $\{\bx\}(t,\btheta)$ is given by
\[
\{ x\}(t;\btheta) = \{X\}(\btheta) + \{\hat u\}(t,\btheta)
\]
with an initial capsule position $\{X\}(\btheta)$ that may depend on $\btheta$ because of the pre-deformation preprocessing
if $a/\ell \geq 0.95$.
}

\begin{figure}
\centering
(a)\includegraphics[width=0.45\textwidth]{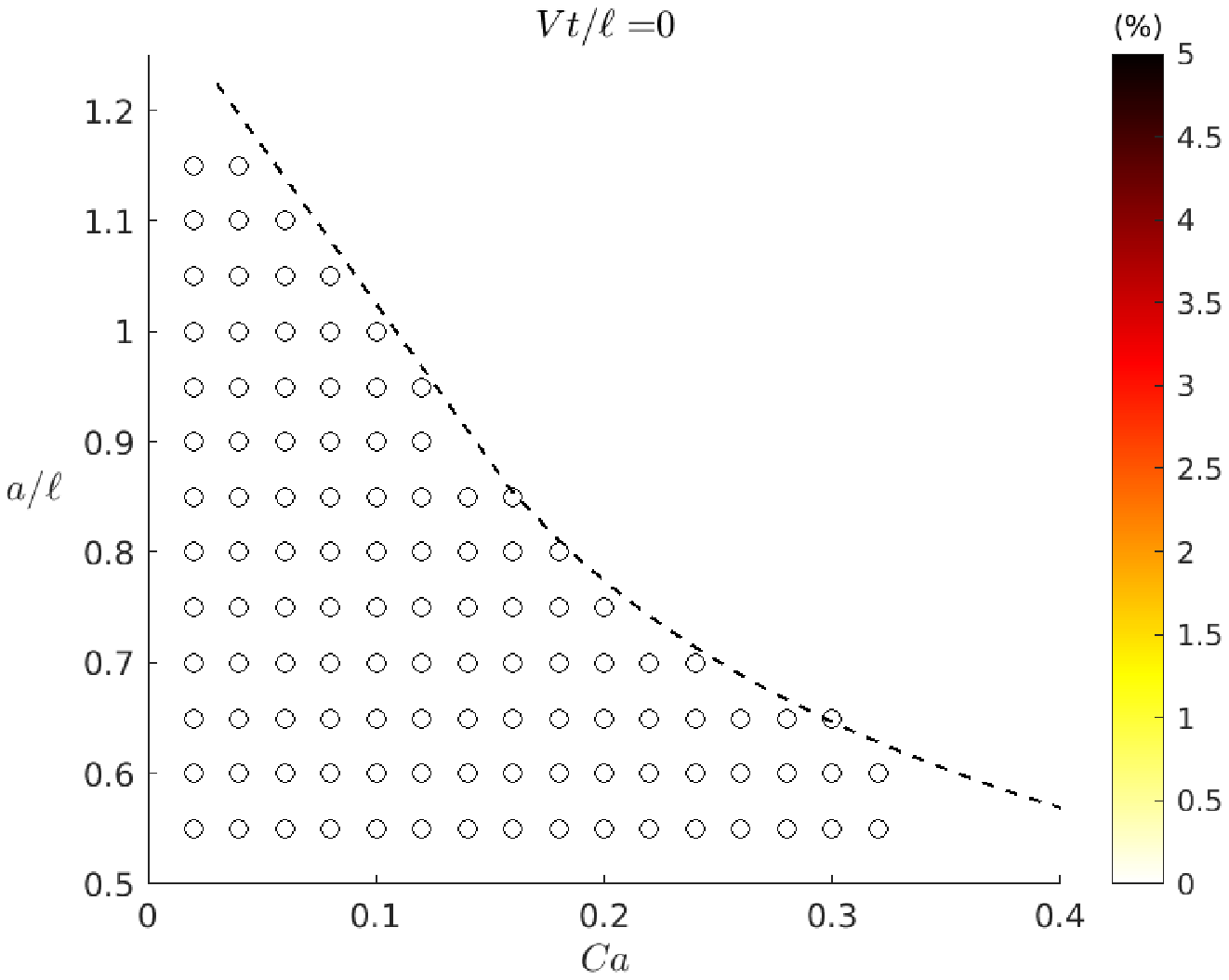}
(b)\includegraphics[width=0.45\textwidth]{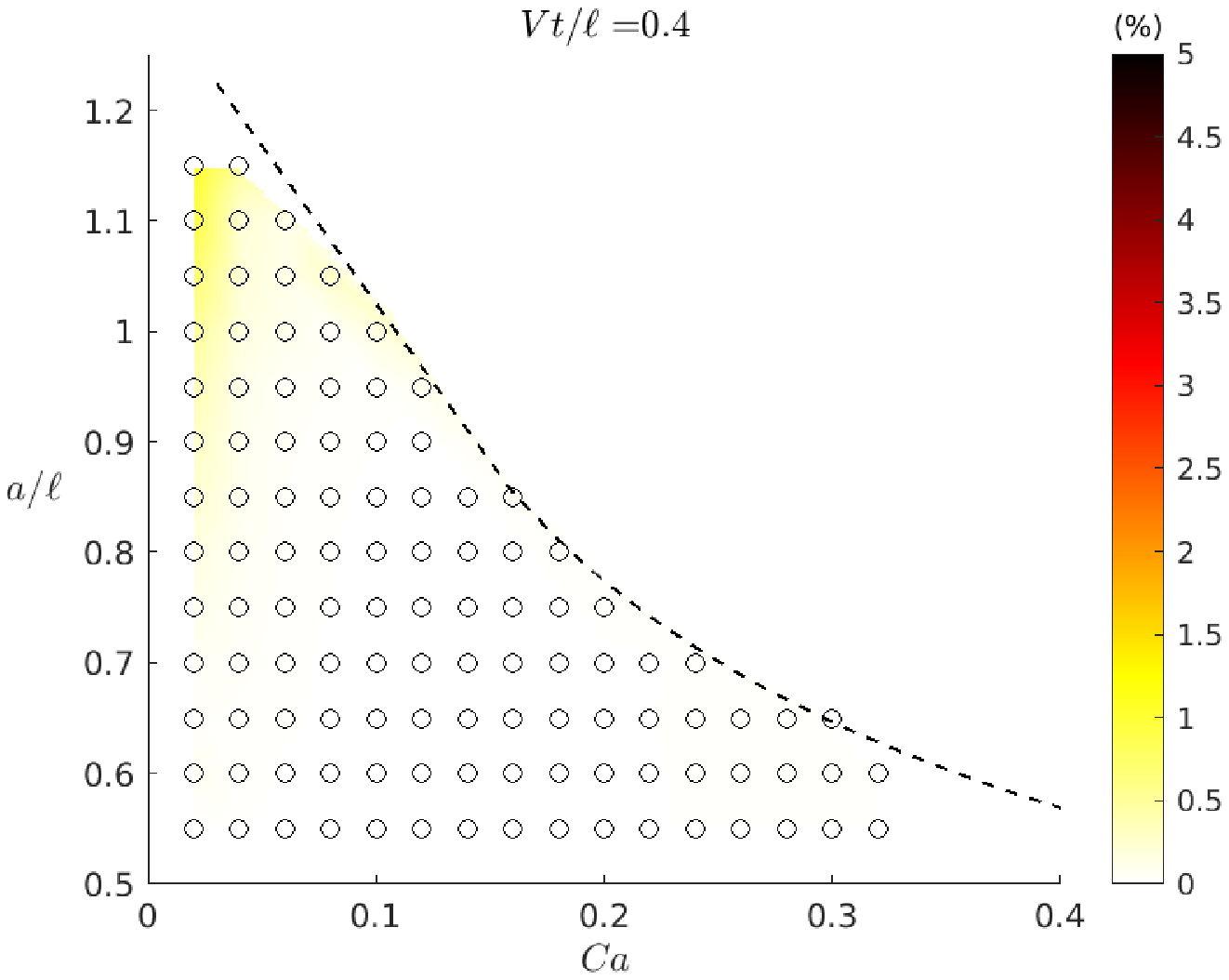}
(c)\includegraphics[width=0.45\textwidth]{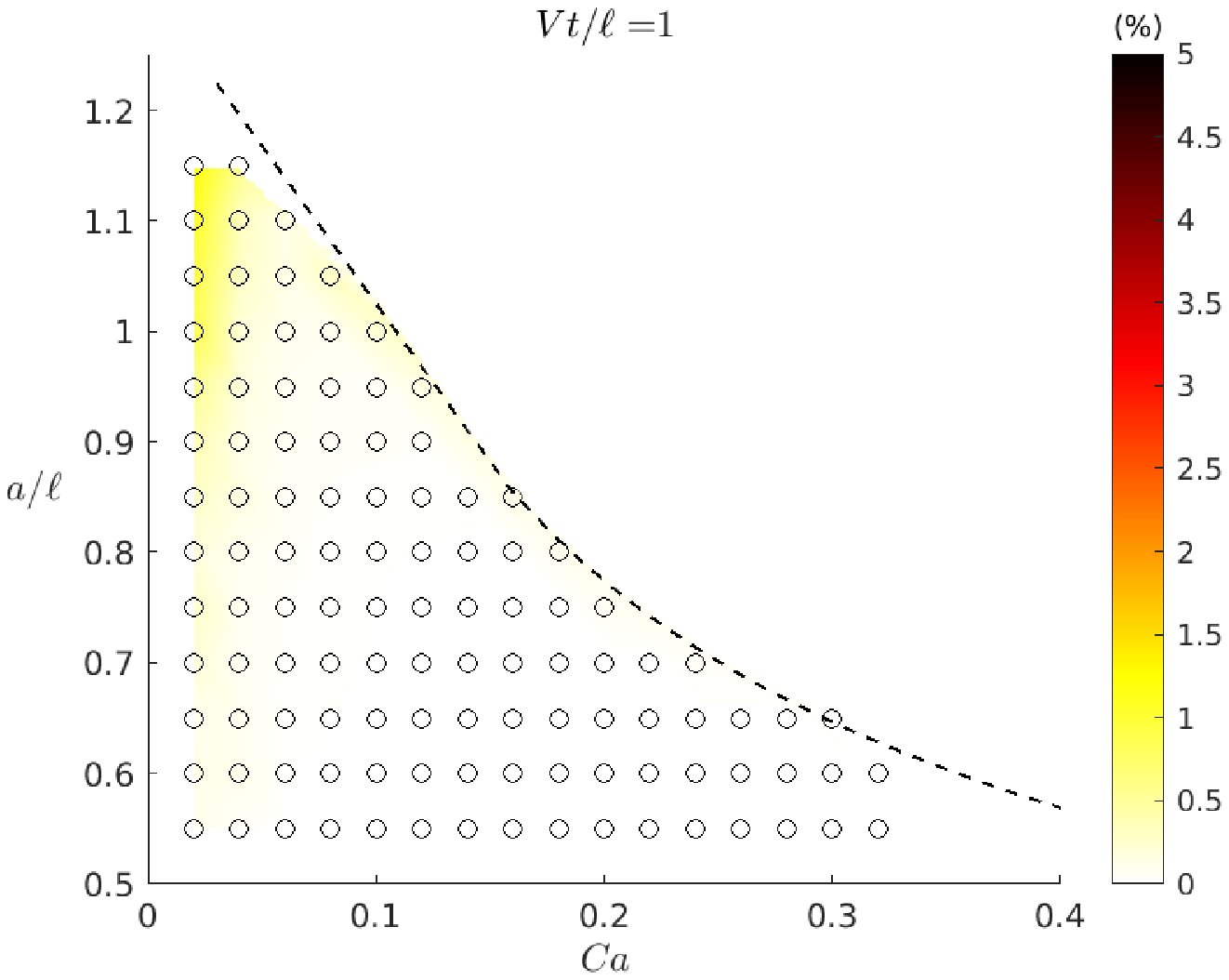}
(d)\includegraphics[width=0.45\textwidth]{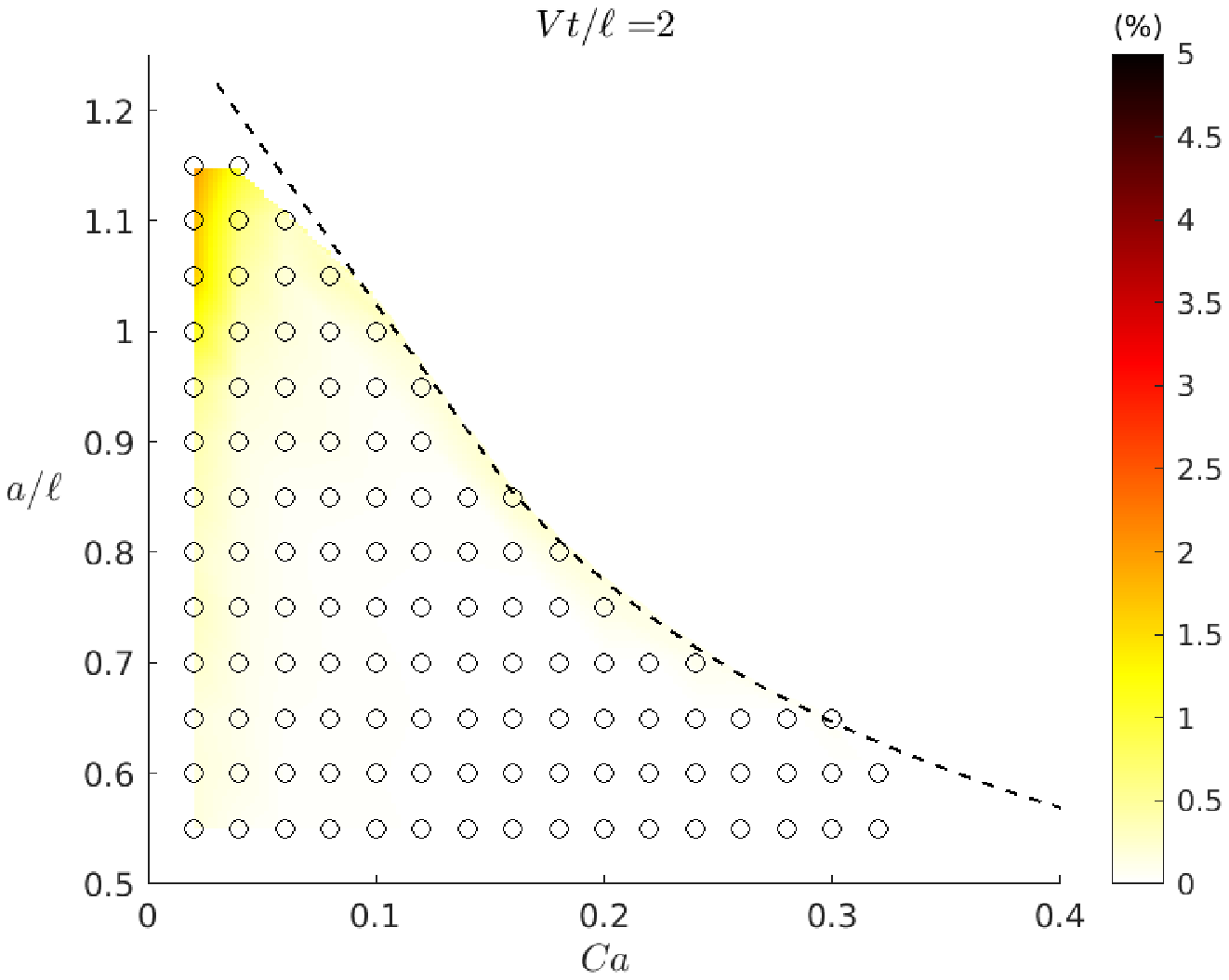}
(e)\includegraphics[width=0.45\textwidth]{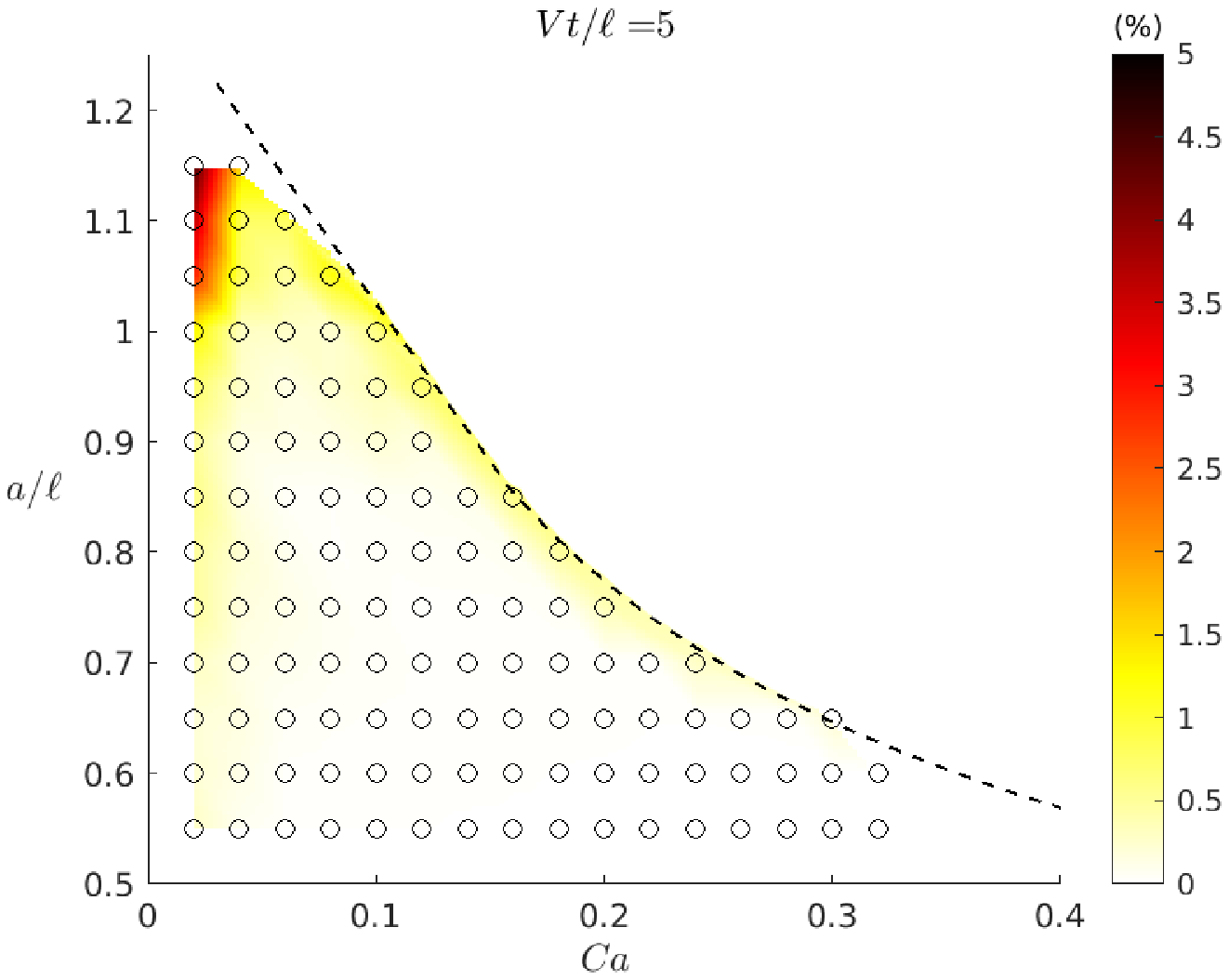}
(f)\includegraphics[width=0.45\textwidth]{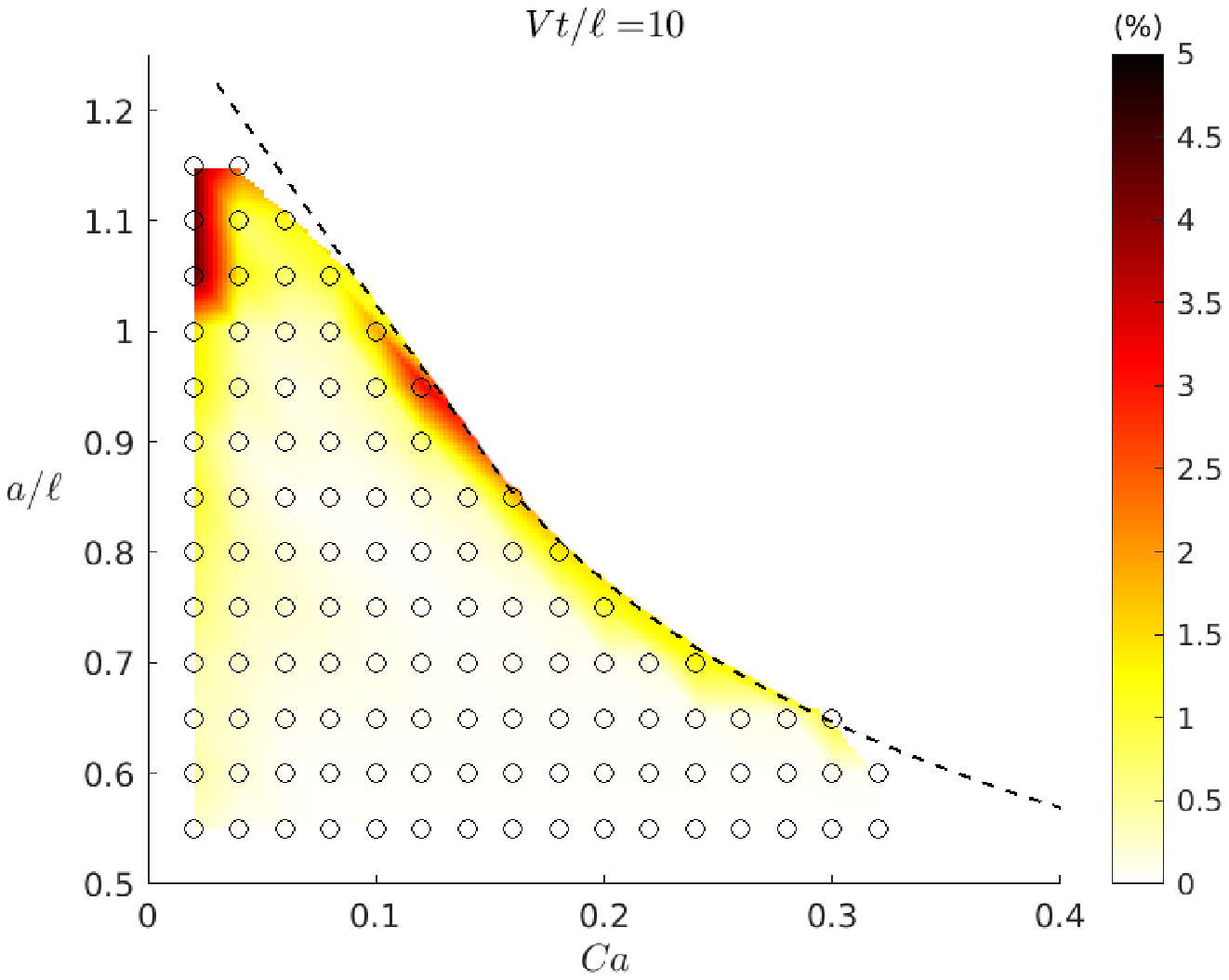}
\caption{\review{Heat maps of $\varepsilon_{\text{Shape}}$ on the testing database as a function of $Ca$ and $a/\ell$ at (a) $\dot{\gamma}t$=0, (b) 0.4, (c) 1, (d) 2, (e) 5, (f) 10. The dotted line delimits the domain for which a steady-state capsule deformation exists.} }
\label{ErreurBaseTestingA20}
\end{figure}

\review{\subsection{Numerical experiments, ROM accuracy assessment}
\begin{figure} 
\centering
\includegraphics[width=0.55\textwidth]{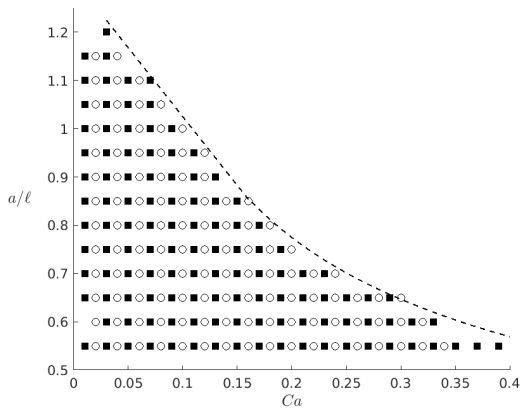}
\caption{Values of $Ca$ and $a/\ell$ included in the testing database (open circle). The filled squares represent the cases in the training database.
The dotted line delimits the domain where a steady-state capsule deformation exists for capsules following the neo-Hookean law.}
\label{DB2}
\end{figure}

A testing database is created using the FOM as in Section \ref{sec:database} and considering $(Ca,a/\ell)$-couples which are not in the training database. A set of 110 $(Ca,a/\ell)$-couples are included in this database (Figure \ref{DB2}). 
For all the $(Ca,a/\ell)$-couples of the testing database, the capsule dynamics is interpolated from the dynamics of the 3 closest neighbors at a given  non-dimensional time. 
Capsule shapes obtained by the ROM are compared to the ones predicted by the FOM at the same nondimensional time. Figure~\ref{ErreurBaseTestingA20} represents the evolution of the error committed on the capsule shape $\varepsilon_{Shape}$ on the training database at $Vt/\ell=0,\ 0.4,\ 1,\ 2,\ 5,\ 10$. At initial time, $\varepsilon_{\text{Shape}}$ is zero. The interpolation method is therefore able to capture the initial capsule shape. When the time increases, $\varepsilon_{\text{Shape}}$ increases and greater than if we apply directly the POD-DMD method on the FOM results and reconstruct the dynamics. However, $\varepsilon_{\text{Shape}}$ remains less than 0.3\% on the majority of the testing database.  It remains fully acceptable.  $\varepsilon_{\text{Shape}}$ is more important near the steady-state limit and when we approach the lowest values of $Ca$ because we are close to the limits of the training base.}

\begin{figure}
\centering
(a)\includegraphics[height=0.17\textwidth]{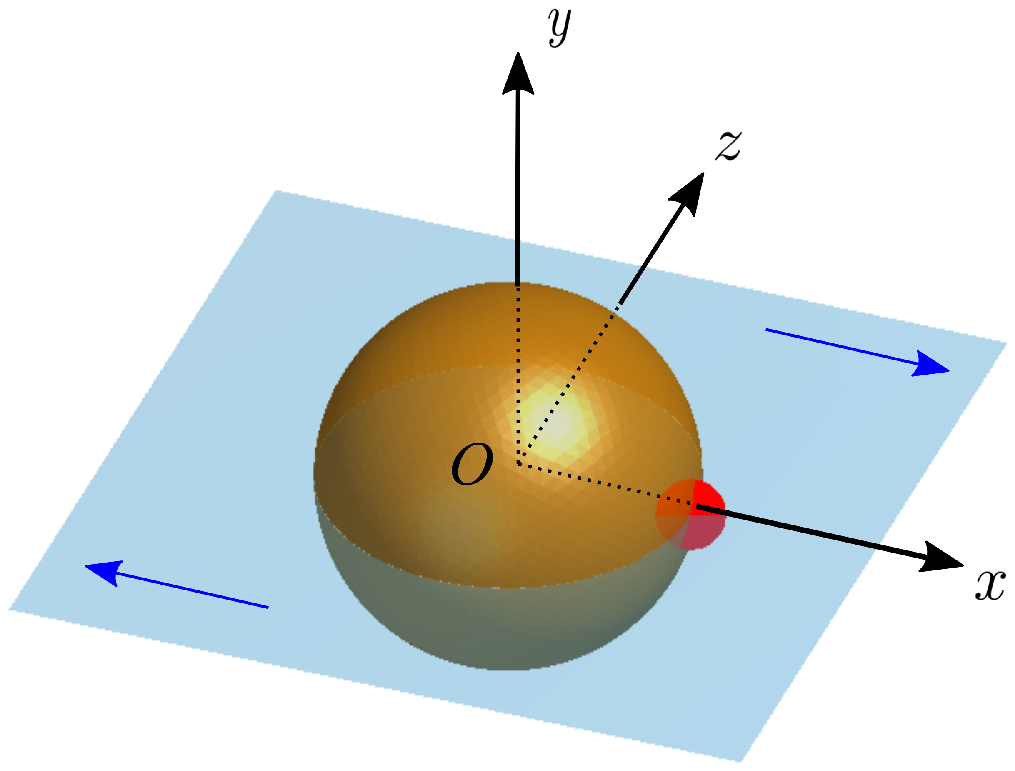}
(b)\includegraphics[height=0.17\textwidth]{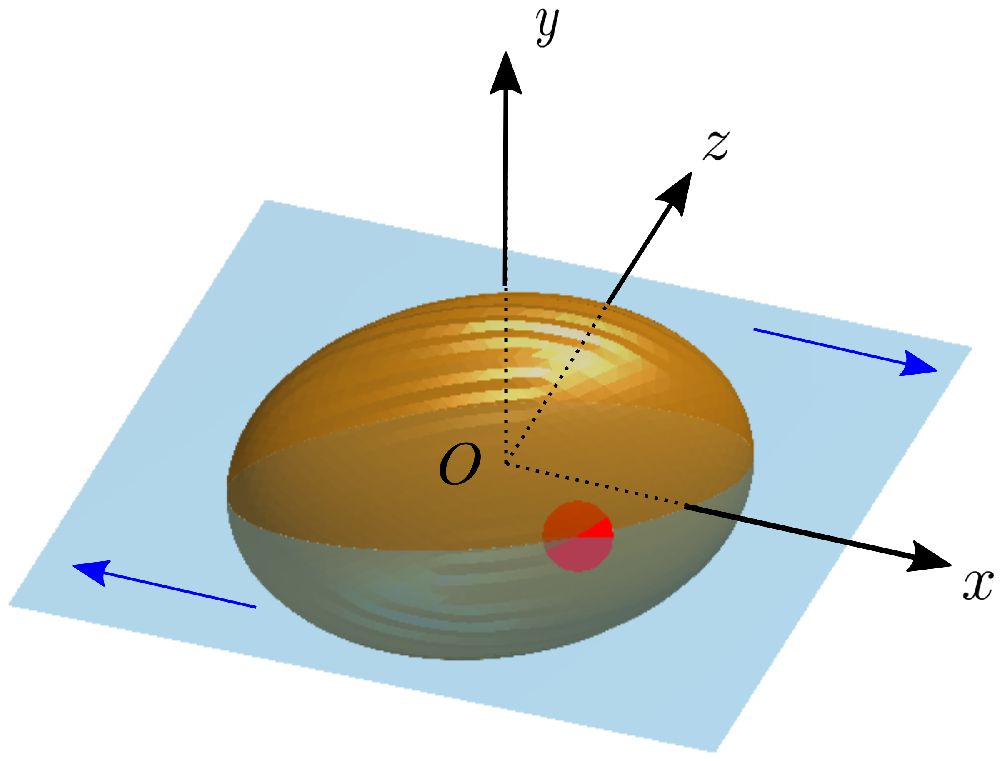}
(c)\includegraphics[height=0.17\textwidth]{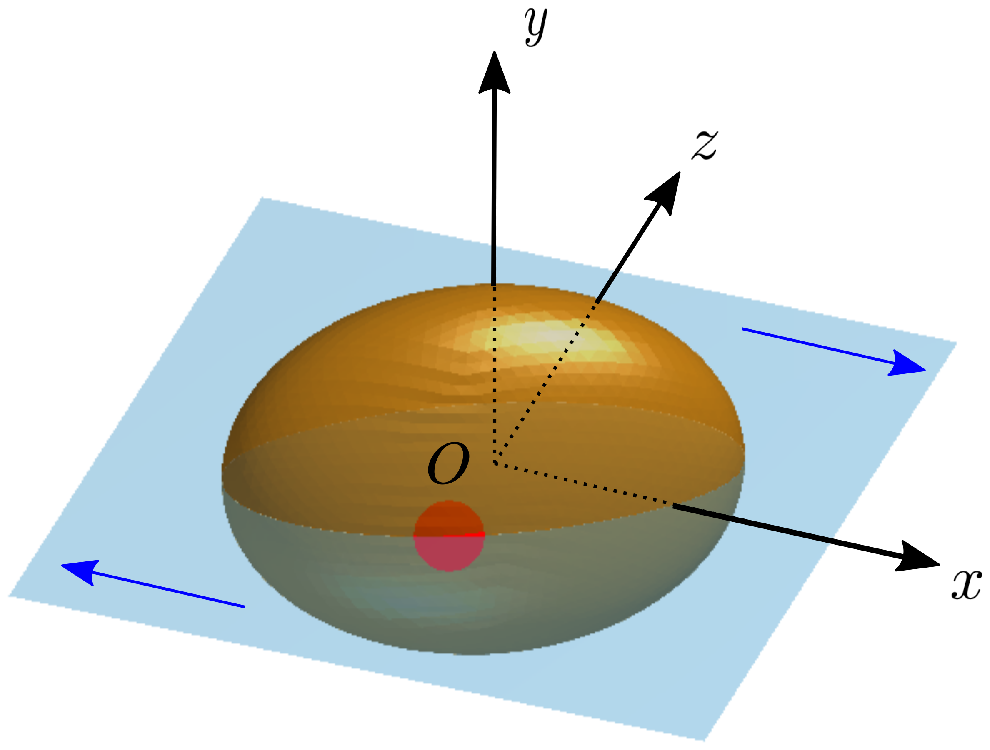}
(d)\includegraphics[height=0.17\textwidth]{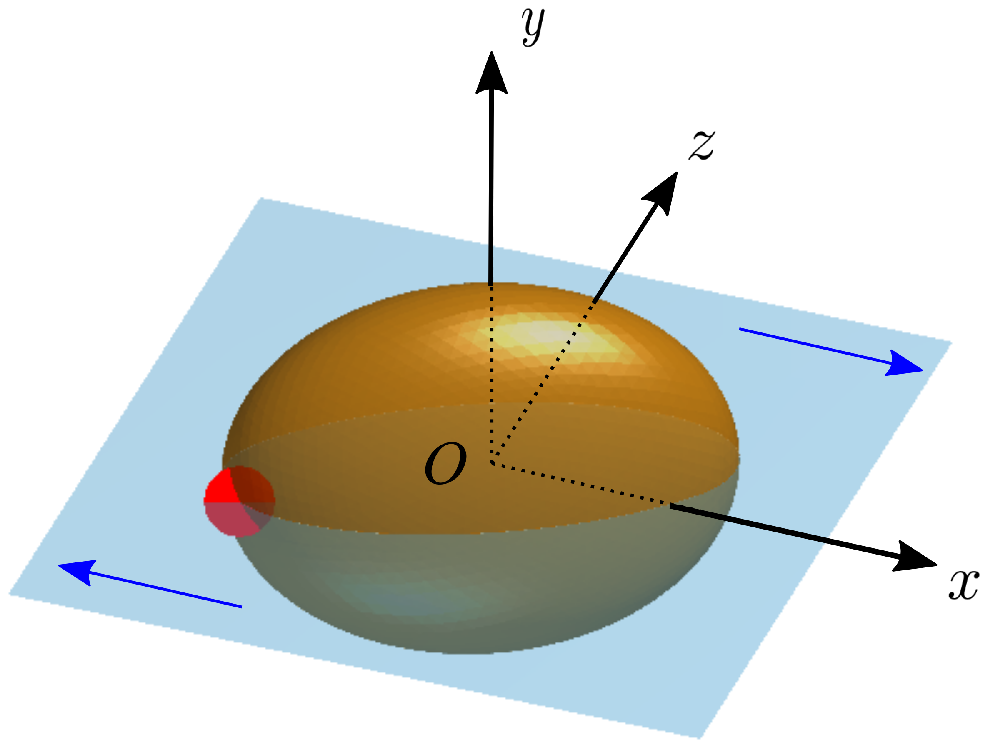}\\
\caption{\review{Dynamics estimated by the ROM of a capsule subjected to a simple shear flow. The capsule is shown for $Ca=0.3$ at the non-dimensional time (a) $\dot{\gamma}=$ 0,(b) 1.6,(c) 4.8,(d) 6.4. 15 modes and $\mu=10^{-6}$ are considered. The red point is a membrane point. }} 
\label{SSF3D}
\end{figure}

\begin{figure}
\centering
\includegraphics[width=\textwidth]{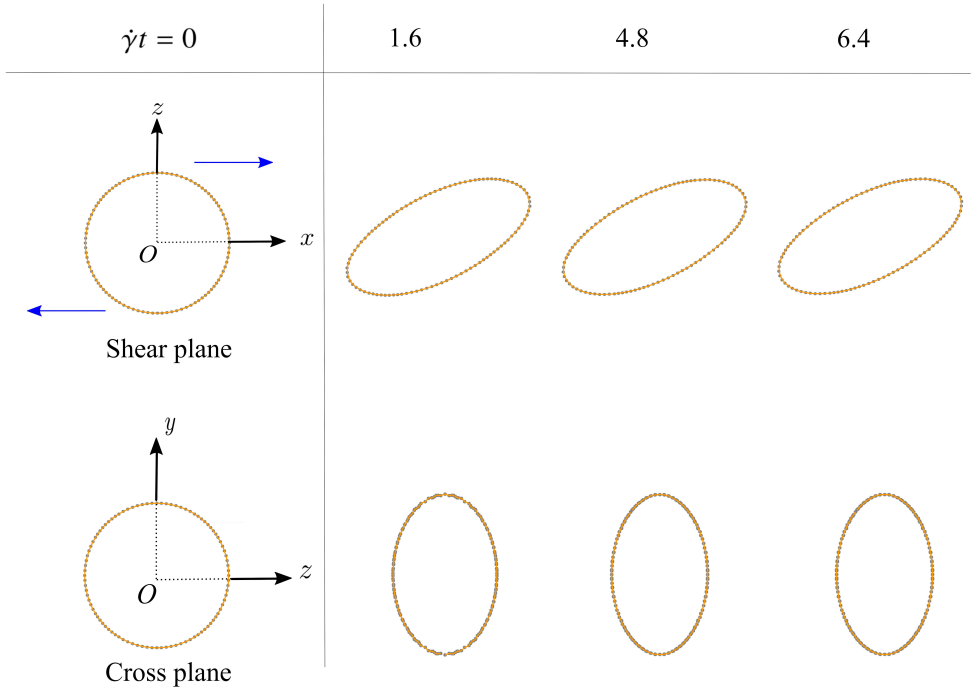}\\
\caption{\review{Capsule subjected to a simple shear flow: Comparison of the  contours given by the FOM (dotted line) and estimated by the ROM (orange line). The capsule is shown for $Ca=0.3$ in the shear plane in the top and in the cross plane in the bottom. 15 modes and $\mu=10^{-6}$ are considered. } } 
\label{SSFCompProfils}
\end{figure}

\begin{figure}
\centering
\includegraphics[width=0.45\textwidth]{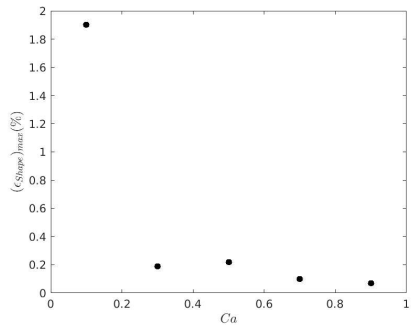}\\
\caption{\review{Evolution of the maximum error committed on the shape of a capsule subjected to a simple shear flow as a function of the capillary number $Ca$. The capsule dynamics was simulated up to a non-dimensional time $\dot{\gamma}=10$.} } 
\label{SSFError}
\end{figure}

\section{\cdd{Discussion and conclusion}} 
\review{
As a summary, in this paper we have considered a $\btheta$-parametrized reduced-order model of microcapsule dynamics in the form
\begin{align*}
    & \dot\balpha(\btheta) = \bbeta(\btheta), \\
    & \dot\bbeta(\btheta) = A_\mu(\btheta)\, \bbeta(\btheta).
\end{align*}

\noindent The vector  $\btheta=(Ca,a/\ell)$ contains the governing parameters, the coefficients $\alpha_k(t,\btheta)$ and $\beta_k(t,\btheta)$ are spectral coefficients of 
POD decomposition for the displacement and velocity fields respectively, and the 
matrix $A_\mu(\btheta)$ is identified from data using a dynamic mode decomposition least-square procedure.
We have numerically proven for a broad range of capillary numbers $Ca$ and aspect ratios $a/\ell$ that it is able to capture the dynamics up to the steady state of a capsule flowing in a channel and its large deformations.
As a first approach, we have presently chosen to use a DMD method that is linear in time to build the ROM model. Still the ROM model captures spatial non-linearity by means of the POD modes. 
The resulting reduced-order model is of great fidelity, weak discrepancies being only observed in the early transient stage.
We have also shown that the learning time need to be larger than the transient stage duration and that we can go beyond the FOM time window used for the training of the ROM model.

For generalisation, we have computed the capsule dynamics for any parameter set. The generalization algorithm is based on interpolation: we first pre-calculate the ROM dynamic model at a finite number of points in the parameter space domain and determine the $\alpha$, $\beta$ and $\phi_k$ (and thus the capsule displacement) at these points. For any other value of the parameters, we first predict the time-evolution of the capsule node displacements using a linear interpolation procedure in the parameter space and then build a dynamical system based the DMD methodology. The error is mostly below 0.3\% over the entire domain, which proves the precision and utility of the ROM approach.

Like any other data-driven model, the model requires a certain number of high-fidelity simulations to provide accurate predictions. By discretizing the parameter space in a regular and homogeneous way (Figure \ref{DB}), we have not presently tried to optimize the number of FOM simulations. But sampling strategies like the Latin Hypercube Sampling (LHS) exist and result in a net reduction in FOM simulation number. 
The empirical law, conventional among the data-driven model community, is that one needs between $10\times D$ and $50 \times D$ points, where $D$ is the dimension of the problem ($D = 2$ in our case). This law shows that the number of high-fidelity simulations does not explode with the problem dimension, owing to the linear dependence of the law.

To prove the generality of the proposed approach, we additionally apply the ROM to a capsule in simple shear flow. This classical case was extensively studied  over the past years \citep{ramanujan98,lac_dbb:05,Li2008,Walter2010,Foessel2011,dbb2010book,dupont2015}.
We build a ROM model to predict the evolution of an initially spherical capsule subjected to a shear rate $\dot{\gamma}$ until $\dot{\gamma}t=10$ with 15 modes, a learning time of $T_L=10$  and $\mu=10^{-6}$. 
The time step~$\Delta t$ between each snapshot is equal to 0.04.
We retrieve that the initial capsule is elongated in the straining direction by the external flow and that the membrane rotates around the deformed shape due to the flow vorticity (Figure~\ref{SSF3D}). The ROM is thus able to recover the tank-treading motion. The capsule contours in both shear and  perpendicular planes predicted by the ROM and simulated by the FOM are in very good agreement (Figure \ref{SSFCompProfils}).
Figure~\ref{SSFError} shows the evolution of the maximum error on the  capsule shape for different values of $Ca$. At $Ca=0.1$, folds appear periodically on the capsule, which prevents the ROM from reproducing precisely the wrinkling phenomenon. 
But from  $Ca\geq 0.3$, the error is reduced by an order of magnitude and is below 0.2\%.

The linear differential model is stable as soon as the eigenvalues of $A_\mu$ have nonpositive real parts, and is consistent with steady states as soon as zero is an eigenvalue. Numerical experiments show that identified matrices $A_\mu$
from data have eigenvalues with negative real parts and one of the eigenvalues is very close to zero.

As it is often the case with spectral-like methods, there is a trade-off between accuracy and ill-conditioning effects: when a large number of POD modes are used ($K>20$), the data matrix~$\mathbb{X}$ of snapshot POD coefficients is ill-conditioned. For the determination of $A_\mu$, we have used a Tikhonov regularization in the least square cost function (see~\eqref{eq:flo21}) in order to have a better conditioned problem and a $L$-curve procedure to determine the best regularization coefficient $\mu$. Unfortunately we observe some limitations in the accuracy. A perspective would be to use a proximal approach: within an iterative procedure, at iteration~$(p+1)$, compute the matrix $A_\mu^{(p+1)}$ solution of
\[
A_\mu^{(p+1)} = \arg 
    \min_{A\in\mathscr{M}_K(\mathbb{R})}\ \frac{1}{2} \|\mathbb{Y}-A\mathbb{X}\|_F^2 + \frac{\mu}{2} \|\mathbb{X}\|_F^2\, \|A-A_\mu^{(p)}\|_F^2
\]
using $A_\mu^{(0)}=0$. At convergence, one can observe that the regularization term vanishes, so that one can expect better accuracy with this approach. This will be investigated in a future work.

We have proposed a  successful and very efficient ROM for FSI problems.
It is an alternative to the use of HPC. It must be seen as a complimentary (and non-competing) approach to full-order models, and has many advantages. 
Among them, one can mention the easiness in implementation. 
It leads to a very handy set of ODEs, that are easy to determine from an algorithmic point of view. Furthermore, the system can be run on any computer. The size of the matrices is, indeed, reduced from ($3 \times 2562 \text{nodes} \times 250 $snapshots) to about ($3 \times 2562 \text{nodes} \times (K+1)$), where the number of modes is $K$ = 20. The computation required time is a few milliseconds for one parameter set. The current  speedups are between 5 000 and 52 000, which out-performs any full-order model approach. We believe that this work is an encouraging milestone to move toward real time simulation of general coupled problems and to deal with high-level parametric studies, sensitivity analysis, optimization and uncertainty quantification.

The next milestone following this work would be to go toward nonlinear differential  dynamical systems as reduced-order models. There is three natural ways for that. The first one is to use Kernel Dynamic Model Decomposition (KDMD) rather than DMD. But we have recently shown in \cite{DeVuyst2022} that a non-linear low-order dynamical model does not provide significant improvement.
The second one is to use Extended Dynamic Model Decomposition (EDMD) \citep{Williams2015}. The EDMD method adds some suitable nonlinear observables (or features) in the data, so that a linear 'augmented' dynamical system is searched for. 
A third option is would be to directly use artificial neural networks (ANN), in particular recurrent neural networks (RNN) \citep{Trischler2016}. The RNN training would replace the DMD procedure, and would be trained with the same POD coefficient matrices $\mathbb{X}$ and~$\mathbb{Y}$. As shown in the recent study by \cite{Lin2021}, artificial intelligent may prove to be efficient and precise to predict capsule deformation.
}

\backsection[Acknowledgements]{
The authors warmly thank Prof. Pierre Villon for fruitful discussions on model-order reduction and related topics.}

\backsection[Funding]{This project has received funding from the European Research Council (ERC) under the European Union’s Horizon
2020 research and innovation programme (Grant agreement No. ERC-2017-COG - MultiphysMicroCaps). }

\backsection[Declaration of interests]{The authors report no conflict of interest.}

\backsection[Author ORCID]{C. Dupont, https://orcid.org/0000-0002-7727-3846; F. De Vuyst, https://orcid.org/0000-0003-0854-4670;  A.-V. Salsac, https://orcid.org/0000-0001-8652-5411}

\backsection[Author contributions]{A.V.S and F.D.V. created the research plan and formulated the numerical problem. C.D. implemented the numerical method and performed the tests.  All authors contributed to analysing data and reaching conclusions, and in writing the paper.}

\appendix
\section{Effects of time derivative discretization on matrix estimation} \label{app:a}
In section~\ref{sec34}, we explain how to identify the coefficient matrix $A$ from a least square problem that tries to minimize the squared residual
$\displaystyle{\int_0^T\|\dot\bbeta(t)-A\, \bbeta(t)\|^2\,dt}$. For practical reasons and because of a finite number of data, we have to discretize the functional and in particular the time derivatives by means of finite differences. This section is dedicated to the analysis of the effect of discretization on the estimation on $A$, and in particular on the effect on the spectrum of $A$ and the impact on the stability of the identified model. \medskip

The notations here are specific to this section. Suppose we have a reference linear dynamical system whose equations and initial data are respectively
\[
\dot \bv = A^{ref}\bv,\ t\in[0,T],\quad \bv(0)=\bv^0\in\mathbb{R}^K,
\]
where $A^{ref}\in\mathscr{M}_K(\mathbb{R})$. The solution of the differential problem problem is given by $\bv(t)=\exp(A^{ref}t)\bv^0$, $t\in[0,T]$. Suppose that we don't know $A^{ref}$ but we have access to the exact solutions
$\bv^n=\bv(t^n)$ at discrete times $t^n=n\Delta t$, $n\in\lbrace 0,...,N\rbrace $ where
with $\Delta t=T/N$. The $(\bv^n)_n$ will be used as data for the  identification (estimation) of the matrix $A^{ref}$. Consider the least square minimization problem
\begin{equation}
\min_{A\in\mathscr{M}_K(\mathbb{R})} \frac{1}{2}\sum_{i=0}^{N-1}
\left\|\frac{\bv^{n+1}-\bv^n}{\Delta t}-A \bv^n\right\|^2.
\label{eq:A1}
\end{equation}
Since $\bv^n=\exp(A^{ref}n\Delta t)\bv^0$ for all $n$, we have also
$\bv^{n+1}-\bv^n = \exp(A^{ref}\Delta t)\bv^n$. So~\eqref{eq:A1} is equivalent to
\[
\min_{A\in\mathscr{M}_K(\mathbb{R})} \frac{1}{2}\sum_{i=0}^{N-1}
\left\| \left( \frac{\exp(A^{ref}\Delta t)-I}{\Delta t}-A \right)\bv^n
\right\|^2 = 
\min_{A\in\mathscr{M}_K(\mathbb{R})} \frac{1}{2}
\left\|
\left( \frac{\exp(A^{ref}\Delta t)-I}{\Delta t}-A \right)\mathbb{X}\right\|_F^2
\]
with $\mathbb{X}=[\bv^0,\bv^1,...,\bv^{N-1}]\in \mathscr{M}_{KN}(\mathbb{R})$.
The first-order optimality conditions are 
\[
A\, \mathbb{X}\mathbb{X}^T = \left(\frac{\exp(A^{ref}\Delta t)-I}{\Delta t}\right)\,\mathbb{X}\mathbb{X}^T.
\]
As soon as $\mathbb{X}$ is a full-rank matrix (meaning that $N\geq K$ and we reasonably have $K$ linearly independent measurements of $\bv^n$), the matrix
$\mathbb{X}\mathbb{X}^T$ is invertible and we get the estimate
\begin{equation}
A = \frac{\exp(A^{ref}\Delta t)-I}{\Delta t}.
\label{eq:A2}
\end{equation}
Let us denote by $\lambda_k^{ref}$ (resp. $\lambda_k$) the (complex)  eigenvalues of $A^{ref}$ (resp. $A$). We have
$\lambda_k = \dfrac{e^{\lambda_k^{ref}\Delta t}-1}{\Delta t}$.
Suppose now that we use a small time step $\Delta t$. From a Taylor expansion,
we observe that
\[
\lambda_k = \lambda_k^{ref}+\frac{\Delta t}{2}(\lambda_k^{ref})^2+o(\Delta t).
\]
We would like to study what is the effect of the first-order error term $\frac{\Delta t}{2}(\lambda_k^{ref})^2$ on the stability of the reconstructed dynamical system $\dot\bv=A\bv$. Suppose that the complex number $\lambda_k^{ref}$ has real and imaginary parts $a$ and $b$ respectively. Then
\[
\lambda_k = \left(a+\frac{\Delta t}{2}(a^2-b^2)\right) + i b(1+a\Delta t)+o(\Delta t).
\]
If $a=\Re(\lambda_k^{ref})\leq 0$, what are the conditions to keep
$\Re(\lambda_k)\leq 0$~? We consider two cases:
\begin{itemize}
    \item If $a=0$ (with $b\neq 0$), $\lambda_k^{ref}$ is pure imaginary, meaning that the $k$-th field is a center for the reference dynamical system. In this case
    $\lambda_k = -\frac{\Delta t}{2}b^2+o(\Delta t)<0$ for a small enough $\Delta t$.
    \item Consider now the case $a\neq 0$. There are two sub-cases. If
    $a^2\leq b^2$, then $\Re(\lambda_k)\leq 0$ for a small enough $\Delta t$.
    If $a^2<b^2$, the condition $\Re(\lambda_k)\leq 0$ gives
    \[
    \Delta t + o(\Delta t) = -\frac{2a}{a^2-b^2}.
    \]
    So there is again a time step $\Delta t^\star>0$ for which, for any $\Delta t<\Delta t^\star$ we have $\Re(\lambda_k)\leq 0$.
\end{itemize}

As a conclusion, starting from a stable linear dynamical system (in the sense that $\Re(\lambda_k^{ref})\leq 0$ for all $k$), using a small enough time step
$\Delta t$ and the forward Euler time discretization, the identification method leads to an estimated dynamical system which is also stable.

Let us underline that this could not be the case using another time discretization as e.g. for the backward Euler time discretization and the associated least square problem
\begin{equation}
\min_{A\in\mathscr{M}_K(\mathbb{R})} \frac{1}{2}\sum_{i=0}^{N-1}
\left\|\frac{\bv^{n+1}-\bv^n}{\Delta t}-A \bv^{n+1}\right\|^2.
\label{eq:A3}
\end{equation}
Using identical developments, we would get in this case
$A=\frac{I-\exp(-A^{ref}\Delta t)}{\Delta t}$ and
\[
\lambda_k = \left(a-\frac{\Delta t}{2}(a^2-b^2)\right) + i b(1-a\Delta t)+o(\Delta t).
\]
We observe that for a center with a pure imaginary eigenvalue $\lambda_k^{ref}=i\,b$, $b\neq 0$, one gets $\lambda_k=\frac{\Delta t}{2}b^2+o(\Delta t)$ therefore $\lambda_k>0$ for a small enough $\Delta t$.
This is a counter-intuitive result: for numerical simulations, it is known that the backward Euler scheme provide more stability than the forward one.
For system identification with time discretization of the residual term, it is safer to use the forward Euler scheme for stability of the estimated dynamical model.

\section{Kinetic energy conservation}
\label{app:b}

    Another quantity of interest is the capsule kinetic energy $\|\vd\|^2$.
    Since the capsules are expected to reach a steady state after a transient stage in the Stokes pipe flow, the kinetic energy should also reach a constant value. From the differential equations, semi-orthogonality of $Q$ and symmetry property of the scalar product, we successively have  
    \begin{align*}
    \frac{d}{dt}\left(\frac{1}{2}\|\vd\|^2\right) & = 
    \frac{d}{dt}\left(\frac{1}{2}\langle Q\bbeta,Q\bbeta \rangle\right) \\
    & = \frac{d}{dt}\left(\frac{1}{2} \|\bbeta\|^2\right) \\
    & = \langle \bbeta,\dot\bbeta\rangle \\
    & = \langle \bbeta, A_\mu \bbeta \rangle \\
    & = \frac{1}{2}\langle \bbeta, A_\mu \bbeta \rangle + \frac{1}{2}
    \langle \bbeta, A_\mu^T \bbeta \rangle \\
    & = \langle \bbeta, \frac{A_\mu+A_\mu^T}{2} \bbeta \rangle.
    \end{align*}
    So stability properties on the kinetic energy are related to the spectral nature of the (symmetric) matrix $A_\mu^S=\frac{A_\mu+A_\mu^T}{2}$. Dissipation property is linked to the non-positiveness of the (real) eigenvalues of $A_\mu^S$.
\section{Practical computation of the pseudo-inverse matrix}
The Moore-Penrose pseudo-inverse $\mathbb{X}^\dagger$ of a matrix $\mathbb{X}$ of
size $d\times K$, $d\geq K$, with $\text{rank}(\mathbb{X})=K$ is defined by
\begin{equation}
\mathbb{X}^\dagger = \mathbb{X}^T (\mathbb{X}\mathbb{X}^T)^{-1}.
\label{app:c1}
\end{equation}
For an ill-conditioned matrix $\mathbb{X}$, the direct computation of $\mathbb{X}^\dagger$ by formula~\eqref{app:c1} is unsuitable because
the condition number of $\mathbb{X}\mathbb{X}^T$ is the square of that of
$\mathbb{X}$. A more robust procedure can be derived by help of the QR
factorization. There exists a semi-orthogonal matrix $\hat Q$ of size $d\times K$
and an upper triangular square matrix $R$ of size $K\times K$ such that
$\mathbb{X}^T = \hat Q R$. Moreover, $R$ is invertible because $\mathbb{X}$ is assumed to be a full-rank matrix. Since $\mathbb{X}^\dagger$ is the solution of the 
matrix system
\[
\mathbb{X}^\dagger \, (\mathbb{X}\mathbb{X}^T)= \mathbb{X}^T,
\]
we get
\[
\mathbb{X}^\dagger \, R^T \hat Q^T \hat Q R = \hat Q R.
\]
By multiplying by $R^{-1}$ to the right, since $\hat Q^T \hat Q=I_K$ we get
\[
\mathbb{X}^\dagger = \hat Q\, (R^T)^{-1} . 
\]

\bibliographystyle{jfm}
\bibliography{CapsuleDMD}
\end{document}